\documentclass[12pt]{article}
\usepackage{amsmath,amssymb,amsfonts,graphicx}

\makeatletter
\@addtoreset{equation}{section}
\makeatother

\begin{document}
\title{Induced quantum numbers of a magnetic vortex \\at nonzero temperature}
\author{Yurii A. Sitenko\thanks{E-mail:
yusitenko@bitp.kiev.ua}\\ \it \small Bogolyubov Institute for
Theoretical Physics, \it \small National Academy of Sciences of
Ukraine,\\ \it \small 14-b Metrologichna str., Kyiv 03143,
Ukraine\\ \phantom{11111111111}\\ Volodymyr M. Gorkavenko
\thanks{E-mail: gorka@univ.kiev.ua}\\
\it \small Department of Physics, Taras Shevchenko National
University of Kyiv,\\ \it \small 6 Academician Glushkov ave., Kyiv
03680, Ukraine\\
\phantom{11111111111}}
\date{}
\maketitle

\begin{abstract}
The phenomenon of the finite-temperature induced quantum numbers
in fermionic systems with topological defects is analyzed. We
consider an ideal gas of twodimensional relativistic massive
electrons in the background of a defect in the form of a pointlike
magnetic vortex with arbitrary flux. This system is found to
acquire, in addition to fermion number, also orbital angular
momentum, spin, and induced magnetic flux, and we determine the
functional dependence of the appropriate thermal averages and
correlations on the temperature, the vortex flux, and the
continuous parameter of the boundary condition at the location of
the defect. We find that nonnegativeness of thermal quadratic
fluctuations imposes a restriction on the admissible range of
values of the boundary parameter. The long-standing problem of the
adequate definition of total angular momentum for the system
considered is resolved.
\end{abstract}
\renewcommand{\theequation}{\arabic{section}.\arabic{equation}}


\section{Introduction}
Quantum fermionic systems in different nontrivial topological
backgrounds (kinks, vortices, monopoles, skyrmeons etc.) can
possess rather unusual properties (e. g. fractionization of
quantum numbers) \cite{1,2}, see reviews in Refs.\cite{3,4}, and
an interest to finite-temperature effects in such systems \cite{5}
has been recently revived \cite{Cor,Du0,Du1,Du2}. In particular,
planar systems with a topological defect in the form of a
pointlike magnetic vortex deserve a thorough examination, since
they may be relevant for the description of some condensed matter
phenomena, including superfluidity  and superconductivity
\cite{Vol, BaE, Fra}, as well as have various applications in
particle physics, cosmology, and astrophysics \cite{Semenr,
Forter}. On the other hand, these systems can be of a certain
conceptual importance, providing a field-theoretical manifestation
of the famous Bohm-Aharonov effect \cite{Aha}: they involve
second-quantized fermions interacting with a vector potential
which is caused by a magnetic flux from the inaccessible for the
fermions region.

A study of quantum numbers which are induced in the Bohm-Aharonov
manner (i. e. by a vector potential of a magnetic vortex) started
in Refs.\cite{Ser,Si8}. It was shown for a particular choice of
the boundary condition at the location of the defect that electric
charge \cite{Si0}, magnetic flux \cite{Gor}, and angular momentum
\cite{SiR} are induced in the vacuum of quantized massive
fermions. The induced vacuum quantum numbers under the most
general set of boundary conditions which are compatible with
self-adjointness of the pertinent Dirac Hamiltonian were obtained
in Refs.\cite{Si6,Si7,Si9}. The finite-temperature induced charge
was examined in Ref.\cite{SiG}. Following this line, we consider
other finite-temperature induced quantum numbers in the present
paper.

We start with the  operator of the second-quantized fermion field
in a static background,
\begin{equation}\label{intr1}
\Psi(\textbf{x},
t)=\sum\hspace{-1.7em}\int\limits_{(E_\lambda>0)}e^{-iE_\lambda
t}\langle\textbf{x}| \lambda\rangle a_{\lambda}+
\sum\hspace{-1.7em}\int\limits_{(E_\lambda<0)}e^{-iE_\lambda
t}\langle\textbf{x}| \lambda\rangle b^+_{\lambda}\,,
\end{equation}
where $a^+_{\lambda}$ and $a_{\lambda}$ $(b^+_{\lambda}$ and
$b_{\lambda})$ are the fermion (antifermion) creation and
destruction  operators satisfying anticommutation relations,
\begin{equation}\label{intr2}
\left[a_{\lambda},a^+_{\lambda'}\right]_+=
\left[b_{\lambda},b^+_{\lambda'}\right]_+= \langle
\lambda|\lambda'\rangle\,,
\end{equation}
and $\langle \textbf{x}|\lambda\rangle$ is the solution to the
stationary Dirac equation,
\begin{equation}\label{intr3}
H\langle \textbf{x}|\lambda\rangle=E_\lambda\langle
\textbf{x}|\lambda\rangle\,,
\end{equation}
$H$ is the Dirac Hamiltonian, $\lambda$ is the set of parameters
(quantum numbers) specifying a one-particle state, and $E_\lambda$
is the energy of the state; symbol
${\displaystyle\sum\hspace{-1.4em}\int\,}$ means the summation
over discrete and the integration (with a certain measure) over
continuous values of $\lambda$.  Ground state $|{\rm vac}\!>$ of
the second-quantized theory is defined as
\begin{equation}\label{intr4}
a_\lambda|{\rm vac}\!>=b_\lambda|{\rm vac}\!>=0\,.
\end{equation}

Let $J$ be an operator commuting with the Hamiltonian in the
first-quantized theory,
\begin{equation}\label{intr5}
  [J,H]_-=0\,.
\end{equation}
In the case of unbounded operators, commutation of their
resolvents is implied, or, to be more specific, it is sufficient
to require that operators $H$ and $J$ have a common set of
eigenfunctions, i.e. relation
\begin{equation}\label{intr6}
J<\textbf{x}|\lambda>=j_\lambda<\textbf{x}|\lambda>
\end{equation}
holds as well as Eq.\eqref{intr3}. Eigenfunctions
$<\textbf{x}|\lambda>$ satisfy the conditions of completeness and
orthonormality; in general, normalization to a delta function is
implied. Thus, in the second-quantized theory, the operators of
the dynamical variables (physical observables) corresponding to
$H$ and $J$ can be diagonalized:
\begin{multline}\label{intr7}
{\hat P}^0\equiv\frac i4\int d^d
x\left\{\left[\Psi^+(\textbf{x},t),\,\partial_t\Psi(\textbf{x},t)\right]_-
-\left[\partial_t\Psi^+(\textbf{x},t),\,\Psi(\textbf{x},t)\right]_-\right\}=\\
=\sum\hspace{-1.3em}\int E_\lambda\left[a^+_\lambda
a_\lambda-b^+_\lambda b_\lambda-\frac12{\rm
sgn}(E_\lambda)\right]\,,
\end{multline}
and
\begin{equation}\label{intr8}
\hat M\equiv\frac12 \int d^d
x\left[\Psi^+(\textbf{x},t),\,J\Psi(\textbf{x},t)\right]_-
=\sum\hspace{-1.3em}\int j_\lambda\left[a^+_\lambda
a_\lambda-b^+_\lambda b_\lambda-\frac12{\rm
sgn}(E_\lambda)\right]\,,
\end{equation}
$d$ is the space dimension.

Let us define partition function
\begin{equation}\label{intr9}
Z(\beta,\mu_J)=Sp\,\exp\left[-\beta\left({\hat P}^0-\mu_J \hat
M\right)\right]\,,\quad \beta=(k_BT)^{-1}\,,
\end{equation}
where $T$ is the equilibrium temperature, $k_B$ is the Boltzmann
constant, $\mu_J$ is the generalized chemical potential, and $Sp$
is the trace or the sum over the expectation values in the Fock
state basis created by operators in Eq.\eqref{intr2}. Although
this sum becomes divergent in the limit of infinite space volume,
this will not bother us, since the partition function plays a
merely supplementary role. The quantities of physical interest are
obtained by taking derivatives of $\ln Z(\beta,\mu_J)$, and these
latter may appear to be finite in the infinite volume limit.

In particular, one can define an average of operator $\hat M$ over
the grand canonical ensemble
\begin{equation}\label{intr10}
\left<\hat
M\right>_{\beta,\mu_J}\equiv\frac1\beta\frac\partial{\partial\mu_J}\ln
Z(\beta,\mu_J)=Z^{-1}(\beta,\mu_J)Sp\,\hat M
\exp\left[-\beta\left({\hat P}^0-\mu_J \hat M\right)\right]\,.
\end{equation}
Computing averages
\begin{equation}\label{intr11}
\begin{array}{lcl}
\left<a^+_\lambda a_\lambda\right>_{\beta,\mu_J}&=&
\{\exp[\beta(E_\lambda-\mu_Jj_\lambda)]+1\}^{-1}\,,\quad
E_\lambda>0\\
 \left<b^+_\lambda b_\lambda\right>_{\beta,\mu_J}&=&
\{\exp[\beta(-E_\lambda+\mu_Jj_\lambda)]+1\}^{-1}\,,\quad
E_\lambda<0
\end{array}\,,
\end{equation}
and using the explicit form of ${\hat P}^0$ and $\hat M$ in terms
of the creation and destruction operators, one gets the spectral
integral representation of average \eqref{intr10} (see, e.g.
Ref.\cite{5}),
\begin{equation}\label{intr12}
\left<\hat
M\right>_{\beta,\mu_J}=-\frac12\int\limits_{-\infty}^{\infty}dE\,\tau_J(E)
\tanh\left[\frac12\beta(E-\mu_Jj)\right]\,,
\end{equation}
where the appropriate spectral density is
\begin{equation}\label{intr13}
\tau_J(E)=\frac1\pi Im\,Tr\,J(H-E-i0)^{-1}\,,
\end{equation}
$Tr$ is the trace of an integro-differential operator in the
functional space: $Tr\,U=\int d^d x\,tr$
$<\textbf{x}|U|\textbf{x}>$; $tr$ denotes the trace over spinor
indices only; note that the functional trace should be regularized
and renormalized by subtraction, if necessary.

Taking $J=I$, where $I$ is the unit matrix in the space of Dirac
matrices, one gets $\hat M=\hat N$, where $\hat N$ is the fermion
number operator in the second-quantized theory, then $\mu_I$ is
the usual chemical potential. In the $d=1$ case fermion number is
the only observable which is conserved in addition to energy. In
more than one dimensions there are more conserved observables. In
particular, in the $d=2$ case, in addition to energy and fermion
number, also total angular momentum is conserved when the system
is rotationally invariant.

Now let us consider an observable which is not conserved and
denote the appropriate operator in the first-quantized theory by
$\Omega$. Then the corresponding operator in the second-quantized
theory,
\begin{equation}\label{intr14}
\hat O=\frac12\int d^d x[\Psi^+,\Omega\Psi]_-\,,
\end{equation}
is not diagonalizable. Nevertheless, its average over the grand
canonical ensemble can be defined in a manner similar to
Eq.\eqref{intr10},
\begin{equation}\label{intr15}
\left<\hat O\right>_{\beta,\mu_J}\equiv
Z^{-1}(\beta,\mu_J)Sp\,\hat O\,\exp\left[-\beta\left({\hat
P}^0-\mu_J\hat M \right)\right]\,.
\end{equation}
One can get an appropriate spectral integral representation,
\begin{equation}\label{intr16}
\left<\hat
O\right>_{\beta,\mu_J}=-\frac12\int\limits_{-\infty}^\infty
dE\,\tau_\Omega(E)\tanh\left[\frac12\beta(E-\mu_Jj)\right]\,,
\end{equation}
where
\begin{equation}\label{intr17}
\tau_\Omega(E)=\frac1\pi Im\, Tr\,\Omega(H-E-i0)^{-1}\,.
\end{equation}

In the present paper we shall be dealing with the averages over
the canonical ensemble:
\begin{equation}\label{intr18}
M(T)\equiv\left<\hat M\right>_{\beta,\mu_J=0}\,,\quad
O(T)\equiv\left<\hat O\right>_{\beta,\mu_J=0}\,.
\end{equation}
In addition to them we shall be considering also such quantities
as the correlation of the conserved and nonconserved observables
\begin{equation}\label{intr19}
\Delta(T;\hat O,\hat M)\equiv\left<{\hat O}{\hat
M}\right>_{\beta,\mu_J=0}-\left<{\hat
O}\right>_{\beta,\mu_J=0}\left<{\hat M}\right>_{\beta,\mu_J=0}\,.
\end{equation}
and the quadratic fluctuation of the conserved observable
\begin{equation}\label{intr20}
\Delta(T;\hat M,\hat M)\equiv\left<{\hat
M}^2\right>_{\beta,\mu_J=0}-\left(\left<\hat
M\right>_{\beta,\mu_J=0}\right)^2\,,
\end{equation}
Using  Eqs.\eqref{intr12} and \eqref{intr16}, one can get the
spectral integral representation for Eqs.\eqref{intr19} and
\eqref{intr20}:
\begin{equation}\label{intr21}
\Delta(T;\hat O,\hat M)=\left.\frac1\beta
\left(\frac{\partial}{\partial\mu_J}\left<{\hat
O}\right>_{\beta,\mu_J}\right)\right|_{\mu_J=0}=\frac14\int\limits_{-\infty}^{\infty}dE\,
\tau_{\Omega J}(E)\,{\rm sech}^2\left(\frac12\beta E\right)\,,
\end{equation}
and
\begin{equation}\label{intr22}
\Delta(T;\hat M,\hat M)=\left.\frac1\beta
\left(\frac{\partial}{\partial\mu_J}\left<{\hat
M}\right>_{\beta,\mu_J}\right)\right|_{\mu_J=0}=\frac14\int\limits_{-\infty}^{\infty}dE\,
\tau_{J^2}(E)\,{\rm sech}^2\left(\frac12\beta E\right)\,,
\end{equation}
where the appropriate spectral densities are obtained from
Eqs.\eqref{intr17} and \eqref{intr13} by inserting an additional
power of operator $J$ into the trace.

\section{Observables of the planar fermionic
system\\ in the background of a magnetic vortex defect}

We consider a spinor field which is quantized in the background of
a static magnetic field in $2+1$-dimensional space-time. The Dirac
Hamiltonian takes form
\begin{equation}\label{b1}
H=-i\,\mbox{\boldmath $\alpha$}\left[\mbox{\boldmath
$\partial$}-ie\textbf{V}(\textbf{x})\right]+\beta m\,,
\end{equation}
where $\textbf{V}(\textbf{x})$ is the vector potential of the
field strength $B(\textbf{x})=\mbox{\boldmath
$\partial$}\times\textbf{V}(\textbf{x})$. The Clifford algebra in
this case has two inequivalent irreducible representations which
can be differed in the following way:
\begin{equation}\label{b2}
\alpha^1\alpha^2\beta=is\,,\quad s=\pm1\,.
\end{equation}
Choosing the $\beta$ matrix in the diagonal form,
\begin{equation}\label{b3}
\beta=\sigma_3\,,
\end{equation}
one gets
\begin{equation}\label{b4}
\alpha^1=-e^{\frac i2 \sigma_3 \chi_s}\sigma_2 e^{-\frac i2
\sigma_3 \chi_s}\,,\quad \alpha^2=se^{\frac i2 \sigma_3
\chi_s}\sigma_1 e^{-\frac i2 \sigma_3 \chi_s}\,,
\end{equation}
where $\sigma_1$, $\sigma_2$, and $\sigma_3$ are the Pauli
matrices, and $\chi_1$ and $\chi_{-1}$ are the parameters varying
in interval $0<\chi_s<2\pi$ to go over to the equivalent
representation. Note also that in odd-dimensional space-time the
$m$ parameter in Eq.\eqref{b1} can take both positive and negative
values; a change of sign of $m$ corresponds to going over to the
inequivalent representation.

If a magnetic field is invariant under rotations of the
twodimensional space around its origin, then one has
\begin{equation}\label{b5}
(\textbf{x}\times\mbox{\boldmath $\partial$})\left[\mbox{\boldmath
$\partial$}\times\textbf{V}(\textbf{x})\right]=0\,,
\end{equation}
and a generator of rotations takes form
\begin{equation}\label{b6}
J=-i\,\textbf{x}\times\left[\mbox{\boldmath
$\partial$}-ie\textbf{V}(\textbf{x})\right]+\frac12 s\beta+e\int\limits_0^r
dr\,r\left[\mbox{\boldmath
$\partial$}\times\textbf{V}(\textbf{x})\right]\,,
\end{equation}
$r=\sqrt{(x^1)^2+(x^2)^2}$ and $\varphi=\arctan(x^2/x^1)$ are the
polar coordinates. One can easily verify that operator
$J$\eqref{b6} commutes with operator $H$\eqref{b1}.

In the right hand side of Eq.\eqref{b6}, the first two terms
represent the orbital and the spin parts of the angular momentum
of the charged matter field, whereas the last term represents the
angular momentum of the background field. In the nonsingular
long-range gauge
\begin{equation}\label{b7}
\textbf{x}\cdot\textbf{V}(\textbf{x})=0\,,
\end{equation}
one gets
\begin{equation}\label{b8}
\textbf{x}\times\textbf{V}(\textbf{x})=\int\limits_0^r
dr\,r\left[\mbox{\boldmath
$\partial$}\times\textbf{V}(\textbf{x})\right]\,,
\end{equation}
and Eq.\eqref{b6} takes form
\begin{equation}\label{b9}
J=-i\,\textbf{x}\times\mbox{\boldmath $\partial$}+\frac12
s\beta\,.
\end{equation}

The above is relevant for the case of an extensive configuration
of the background magnetic field (see, e.g., Ref.\cite{Paranr}).
Turning now to the case of the background in the form of a
magnetic vortex defect, let the central region (e.g. a disc of
radius $\delta$) be impenetrable for the charged matter and the
background field strength be nonvanishing only in this region
(i.e. the region of the defect). Then the angular momentum
operator outside the central region consists of two parts, orbital
and spin,
\begin{equation}\label{b10}
J=-i\,\textbf{x}\times\left[\mbox{\boldmath
$\partial$}-ie\textbf{V}(\textbf{x})\right]+\frac12 s\beta\,.
\end{equation}
As is well known (see, e.g., Ref.\cite{Aha}), due to nonvanishing
flux of the background field in the inner region,
\begin{equation}\label{b11}
\Phi=\int\limits_0^\delta
dr\,r\left[\mbox{\boldmath
$\partial$}\times\textbf{V}(\textbf{x})\right]\,,
\end{equation}
the vector potential cannot be made vanishing everywhere in the
outer region. In particular, in the gauge \eqref{b7} one gets at
$r>\delta$:
\begin{equation}\label{b12}
V^1(\textbf{x})=-\Phi r^{-1}\sin\varphi\,,\quad
V^2(\textbf{x})=\Phi r^{-1}\cos\varphi\,,
\end{equation}
and Eq.\eqref{b10} takes form
\begin{equation}\label{b13}
J=-i\,\textbf{x}\times\mbox{\boldmath $\partial$}-e\Phi+\frac12
s\beta\,.
\end{equation}

Thus, contrary to the case of the extensive background field
configuration when the angular momentum is quantized in
half-integer values,
\begin{equation}\label{b14}
j=n+\frac12\,,
\end{equation}
(this is evident in the gauge \eqref{b7}, see Eq.\eqref{b9}), in
the case of the vortex defect, the angular momentum is quantized
in units
\begin{equation}\label{b15}
j=n+\frac12-e\Phi\,,
\end{equation}
and this results in such fascinating quantum-mechanical concepts
as anyons and fractional statistics \cite{Wil82}. However, in the
latter case one can take operator
\begin{equation}\label{b16}
J'=-i\,\textbf{x}\times\left[\mbox{\boldmath
$\partial$}-ie\textbf{V}(\textbf{x})\right]+\frac12 s\beta+\Xi\,,
\end{equation}
as well as an operator of conserved quantity; here $\Xi$ is an
arbitrary constant. In particular, choosing $\Xi=e\Phi$, one gets
in the gauge \eqref{b7} the same expression as Eq.\eqref{b9} and,
consequently, half-integer eigenvalues. The arguments in favour of
such a definition of the angular momentum operator are given in
Ref.\cite{Jac83}. Not going into details of the discussion at the
quantum-mechanical level, we would like to emphasize that the
problem of the proper definition of the angular momentum operator
might be resolved in the framework of the second-quantized theory
at nonzero temperature. Indeed, the quadratic fluctuation
\eqref{intr22} of the physically meaningful observable has to be
nonnegative, and this, as we shall see in Section 6, imposes a
definite restriction on the choice of the appropriate operator in
the first-quantized theory.

Turning to the nonconserved observables, it is natural to
consider, in the capacity of $\Omega$, the orbital angular
momentum operator,
\begin{equation}\label{b17}
\Lambda=-i\,\textbf{x}\times\left[\mbox{\boldmath
$\partial$}-ie\textbf{V}(\textbf{x})\right]\,,
\end{equation}
and the spin operator,
\begin{equation}\label{b18}
\Sigma=\frac12 s\beta\,.
\end{equation}
In addition to these, we consider also operator \footnote{Note
that $e^2$ has dimension of mass in the case of $2+1$-dimensional
space-time. Hence $\Omega$\eqref{b19} is dimensionless, as well as
$\Lambda$\eqref{b17} and $\Sigma$\eqref{b18}.}
\begin{equation}\label{b19}
\Omega=\frac{e^2}{4\pi}\,\textbf{x}\times\mbox{\boldmath
$\alpha$}\,,
\end{equation}
which corresponds to the observable with the physical meaning of
the induced magnetic flux multiplied by $e$. Really, the latter
quantity at finite temperature is
\begin{equation}\label{b20}
O(T)=e\int\limits_\delta^\infty dr\,r B^{(I)}(r)\,,
\end{equation}
where the induced magnetic field strength $B^{(I)}$ is
rotationally invariant and satisfies Maxwell equation
\begin{equation}\label{b21}
\partial_r\left[r
B^{(I)}(r)\right]=\textbf{x}\times\textbf{j(x)}\,,
\end{equation}
with induced current
\begin{equation}\label{b22}
\textbf{j(x)}=-\frac e2 tr\left<\textbf{x}\left|\mbox{\boldmath
$\alpha$}\tanh\left(\frac12 \beta
H\right)\right|\textbf{x}\right>\,.
\end{equation}
Solving Eq.\eqref{b21} and substituting the solution into
Eq.\eqref{b20}, one gets
\begin{equation}\label{b23}
O(T)=\frac{e}{4\pi}\int\limits_{r>\delta}d^2x\,[\textbf{x}\times\textbf{j(x)}]\,,
\end{equation}
which is the thermal average of operator $\hat O$\eqref{intr14}
constructed from $\Omega$\eqref{b19}.

In the present paper we shall compute thermal averages of the
above observables, as well as correlations of conserved and
nonconserved observables and quadratic fluctuations of conserved
observables; note that the average and fluctuation of charge (i.
e. fermion number times $e$) have been computed earlier
\cite{SiG}. Thermal characteristics of the quantized fermionic
matter in the background of a magnetic vortex defect depend both
on vortex flux \eqref{b11} and the parameter of the boundary
condition for the matter field at the edge of the defect.


\section{Spectral densities and traces of resolvents}

In order to compute thermal characteristics one has to determine
spectral densities $\tau_\Lambda(E)$, $\tau_\Sigma(E)$,
$\tau_\Omega(E)$, $\tau_{\Sigma J}(E)$, $\tau_{J^2}(E)$,
$\tau_{\Omega J}(E)$, which are imaginary parts of the appropriate
functional traces, see, e.g., Eqs.\eqref{intr13} and
\eqref{intr17}. Actually, integrals over the real energy spectrum
can be transformed into integrals over a contour on the complex
energy plane, thus yielding a representation of thermal
characteristics through the traces directly. In particular, one
gets
\begin{equation}\label{c1}
M(T)=-\frac12\int\limits_C\frac{d\omega}{2\pi i}\tanh\left(\frac12
\beta\omega \right)Tr\,J(H-\omega)^{-1}\,,
\end{equation}
and
\begin{equation}\label{c2}
\Delta(T;\hat M,\hat M)=\frac14 \int\limits_C\frac{d\omega}{2\pi
i}\,{\rm sech}^2\left(\frac12 \beta\omega
\right)Tr\,J^2(H-\omega)^{-1}\,,
\end{equation}
and similarly for other characteristics; here $C$ is the contour
$(-\infty+i0,+\infty+i0)$ and $(+\infty-i0,-\infty-i0)$ in the
complex $\omega$-plane.

The kernel of the resolvent (the Green's function) of the Dirac
Hamiltonian in the coordinate representation is defined as
\begin{equation}\label{c3}
G^\omega(r,\varphi;r',\varphi')=\langle
r,\varphi|(H-\omega)^{-1}|r',\varphi' \rangle\,.
\end{equation}
Using Eqs.\eqref{b3} and \eqref{b4}, one can expand Eq.\eqref{c3}
in modes in the following form
\begin{equation}\label{c4}
G^\omega(r,\varphi;r',\varphi')=\frac1{2\pi}\sum_{n\in \mathbb Z}
e^{in(\varphi-\varphi')} \left(\begin{array}{cc} a_n(r;r')&
d_n(r;r')e^{-i(s\varphi'-\chi_s)}\\
b_n(r;r')e^{i(s\varphi-\chi_s)} &
c_n(r;r')e^{is(\varphi-\varphi')}\end{array}\right).
\end{equation}
In the background of magnetic vortex defect \eqref{b12},
Hamiltonian \eqref{b1} takes form
\begin{equation}\label{c5}
H=-i\alpha^r\partial_r-ir^{-1}\alpha^\varphi
(\partial_\varphi-ie\Phi)+\beta m\,,
\end{equation}
where
\begin{equation}\label{c6}
\alpha^r=\alpha^1\cos\varphi+\alpha^2
\sin\varphi\,,\quad\alpha^\varphi=-\alpha^1\sin\varphi+\alpha^2\cos\varphi\,.
\end{equation}
If a size of the defect is neglected $(\delta\rightarrow0)$, then
a parameter of the boundary condition at the location of the
defect (at $r=0$) exhibits itself as a parameter of a self-adjoint
extension of the Hamiltonian operator. Partial Hamiltonians are
essentially self-adjoint for all $n$, with the exception of
$n=n_0$, where
\begin{equation}\label{c7}
n_0=[\![e\Phi]\!]+\frac12-\frac12s\,,
\end{equation}
$[\![u]\!]$ is the integer part of quantity $u$ (i.e., the largest
integer which is less than or equal to $u$).  The partial
Hamiltonian for $n=n_0$ requires a self-adjoint extension
according to the Weyl-von Neumann theory of self-adjoint operators
(see, e.g., Ref.\cite{Alb}). Appropriately, radial components
$a_n$, $b_n$, $c_n$, and $d_n$ in Eq.\eqref{c4} with $n\neq n_0$
are regular at $r\rightarrow0$ and $r'\rightarrow0$, whereas those
with $n=n_0$ satisfy conditions (for details see Ref.\cite{SiG}):
\begin{equation}\label{c8}
\left.\begin{array}{l}
{\displaystyle\cos\left(s\frac\Theta2+\frac\pi4\right)\lim_{r\rightarrow0}(|m|r)^Fa_{n_0}(r;r')}=
{\displaystyle-{\rm{sgn}}(m)\sin\left(s\frac\Theta2+\frac\pi4\right)\lim_{r\rightarrow0}(|m|r)^{1-F}b_{n_0}(r;r')}\vspace{0.3em}\\
{\displaystyle\cos\left(s\frac\Theta2+\frac\pi4\right)\lim_{r\rightarrow0}(|m|r)^Fd_{n_0}(r;r')}=
{\displaystyle-{\rm{sgn}}(m)\sin\left(s\frac\Theta2+\frac\pi4\right)\lim_{r\rightarrow0}(|m|r)^{1-F}c_{n_0}(r;r')}
\end{array}\right\}\,,
\end{equation}
and
\begin{equation}\label{c9}
\left.\begin{array}{l}
{\displaystyle\cos\left(s\frac\Theta2+\frac\pi4\right)\lim_{r'\rightarrow0}(|m|r')^Fa_{n_0}(r;r')}=
{\displaystyle-{\rm{sgn}}(m)\sin\left(s\frac\Theta2+\frac\pi4\right)\lim_{r'\rightarrow0}(|m|r')^{1-F}d_{n_0}(r;r')}\vspace{0.3em}\\
{\displaystyle\cos\left(s\frac\Theta2+\frac\pi4\right)\lim_{r'\rightarrow0}(|m|r')^Fb_{n_0}(r;r')}=
{\displaystyle-{\rm{sgn}}(m)\sin\left(s\frac\Theta2+\frac\pi4\right)\lim_{r'\rightarrow0}(|m|r')^{1-F}c_{n_0}(r;r')}
\end{array}\right\}\,,
\end{equation}
where
\begin{equation*}
{\rm{sgn}}(u)= \left\{\begin{array}{cc} 1\,,& u>0\\ -1\,,&
u<0\end{array}\right\}\,,
\end{equation*}
$\Theta$ is the self-adjoint extension parameter, and
\begin{equation}\label{c10}
F=s\{\![e\Phi]\!\}+\frac12-\frac12s\,,
\end{equation}
$\{\![u]\!\}=u-[\![u]\!]$ is the fractional part of quantity $u$,
$0\leq \{\![u]\!\} <1$; note here that Eqs.\eqref{c8} and
\eqref{c9} imply that $0<F<1$, since in the case of
${\displaystyle F=\frac12-\frac12s}$ all radial components obey
the condition of regularity at $r\rightarrow0$ and
$r'\rightarrow0$. Note also that Eqs.\eqref{c8} and \eqref{c9} are
periodic in $\Theta$ with period $2\pi$.

The radial components of the resolvent kernel have been determined
in Ref.\cite{SiG}, and we list them in Appendix A.

Let us consider quantities
\begin{equation}\label{c11}
\int\limits_0^\infty d\varphi\,tr\left[\Lambda\,
G^\omega(r,\varphi;r',\varphi)\right]=\sum_{n=-\infty}^\infty[(n-e\Phi)a_n(r;r')+(n+s-e\Phi)c_n(r;r')]\,,
\end{equation}
\begin{equation}\label{c12}
\int\limits_0^\infty d\varphi\,tr\left[\Sigma\,
G^\omega(r,\varphi;r',\varphi)\right]=\frac12 s
\sum_{n=-\infty}^\infty[a_n(r;r')-c_n(r;r')]\,,
\end{equation}
\begin{equation}\label{c13}
\int\limits_0^\infty d\varphi\,tr\left[\Omega\,
G^\omega(r,\varphi;r',\varphi)\right]=\frac{e^2}{4\pi}sr\sum_{n=-\infty}^\infty[b_n(r;r')+d_n(r;r')]\,,
\end{equation}
\begin{equation}\label{c14}
\int\limits_0^\infty d\varphi\,tr\left[\Sigma J\,
G^\omega(r,\varphi;r',\varphi)\right]=\frac12
s\sum_{n=-\infty}^\infty\left(n-e\Phi+\frac12
s\right)[a_n(r;r')-c_n(r;r')]\,,
\end{equation}
\begin{equation}\label{c15}
\int\limits_0^\infty d\varphi\,tr\left[J^2\,
G^\omega(r,\varphi;r',\varphi)\right]=\sum_{n=-\infty}^\infty\left(n-e\Phi+\frac12
s\right)^2[a_n(r;r')+c_n(r;r')]\,,
\end{equation}
\begin{equation}\label{c16}
\int\limits_0^\infty d\varphi\,tr\left[\Omega J\,
G^\omega(r,\varphi;r',\varphi)\right]=\frac{e^2}{4\pi}sr
\sum_{n=-\infty}^\infty\left(n-e\Phi+\frac12
s\right)[b_n(r;r')+d_n(r;r')]\,,
\end{equation}
where operators $J$, $\Lambda$, $\Sigma$, $\Omega$ are given by
Eqs.\eqref{b13}, \eqref{b17}-\eqref{b19}, correspondingly. Using
the explicit form of $a_n$, $b_n$, $c_n$, $d_n$ given in Appendix
A, we perform summation over $n$ and get in the case of $r'>r$ and
$Im\,k>|Re\, k |$ (see Appendix B):
\begin{multline}\label{c17}
\int\limits_0^{2\pi} d\varphi\,tr\left[\Lambda\,
G^\omega(r,\varphi;r',\varphi)\right]=\frac{s\sin(F\pi)}{\pi}\,\omega\!\!\int\limits_0^\infty\!\!
dy\,\exp\!\left(\!\!-\frac{\kappa^2r r'}{2y}-\frac{r^2+{r'}^2}{2r
r'}\,y\!\right)\![K_F(y)-K_{1-F}(y)]+\\+\frac{2s\sin(F\pi)}{\pi(\tan\nu_\omega+e^{i
F\pi})}\!\left[\!-F(\omega\!+\!m)\tan\nu_\omega K_F(\kappa
r)K_F(\kappa r')\!+ \!(1\!-\!F)(\omega\!-\!m)e^{i
F\pi}K_{1-F}(\kappa r)K_{1-F}(\kappa r')\right],
\end{multline}
\begin{multline}\label{c18}
\int\limits_0^{2\pi} \!\!d\varphi\,tr\left[\Sigma\,
G^\omega(r,\varphi;r',\varphi)\right]\!=\!
sm\,K_0(\kappa|r-r'|)-\frac{s\sin(F\pi)}{4F(1-F)\pi
}\,m\!\!\int\limits_0^\infty\!\! dy\,\exp\left(\!-\frac{\kappa^2r
r'}{2y}-\frac{r^2+{r'}^2}{2r r' }\,y\!\right)\!\!\!\times\\ \times
\left\{\frac{e^y}{y}\int\limits_y^\infty du\,e^{-u}[(1-F)K_F(u)+F
K_{1-F}(u)]-(2F-1)[K_F(y)-K_{1-F}(y)]\right\}+\\
+\frac{s\sin(F\pi)}{\pi(\tan\nu_\omega+e^{i
F\pi})}\!\left[(\omega\!+\!m)\tan\nu_\omega K_F(\kappa
r)K_F(\kappa r')- (\omega-m)e^{i F\pi}K_{1-F}(\kappa
r)K_{1-F}(\kappa r')\right],
\end{multline}
\begin{multline}\label{c19}
\int\limits_0^{2\pi} d\varphi\,tr\left[\Omega\,
G^\omega(r,\varphi;r',\varphi)\right]=\!\frac{e^2s\sin(F\pi)}{4\pi^2}\!\!\int\limits_0^\infty\!\!
dy\,\exp\!\left(\!\!-\frac{\kappa^2r r'}{2y}-\frac{r^2+{r'}^2}{2r
r'}\,y\!\right)[K_F(y)-K_{1-F}(y)]+\\+\frac{e^2s\sin(F\pi)}{2\pi^2(\tan\nu_\omega+e^{i
F\pi})}\,\kappa r\left[\tan\nu_\omega K_{1-F}(\kappa r)K_F(\kappa
r')-e^{i F\pi}K_{F}(\kappa r)K_{1-F}(\kappa r')\right]\,,
\end{multline}
\begin{multline}\label{c20}
\int\limits_0^{2\pi} \!\!d\varphi\,tr\left[\Sigma J\,
G^\omega(r,\varphi;r',\varphi)\right]\!=\!
\frac\omega2\,K_0(\kappa|r-r'|)
-\frac{\sin(F\pi)}{2\pi}\int\limits_0^\infty
dy\,\exp\left(-\frac{\kappa^2r r'}{2y}-\frac{r^2+{r'}^2}{2r r'
}\,y\right)\times\\
\times\left\{\!\frac{\omega}{4F(1-F)}\frac{e^y}{y}\!\int\limits_y^\infty\!\!
du\,e^{-u}[(1\!-\!F)K_F(u)\!+\!F
K_{1-F}(u)]\!-\!\!\left[\frac{\left(F-\frac12\right)\omega}{2F(1-F)}+m\right]\![K_F(y)\!-\!K_{1-F}(y)]\!\right\}\!\!-\\
-\frac{\left(F-\frac12\right)\sin(F\pi)} {\pi(\tan\nu_\omega+e^{i
F\pi})}\!\left[(\omega\!+\!m)\tan\nu_\omega K_F(\kappa
r)K_F(\kappa r')-(\omega-m)e^{i F\pi}K_{1-F}(\kappa
r)K_{1-F}(\kappa r')\right],
\end{multline}
\begin{multline}\label{c21}
\int\limits_0^{2\pi} d\varphi\,tr\left[J^2\,
G^\omega(r,\varphi;r',\varphi)\right]=\frac\omega2\left[K_0(\kappa|r-r'|)+4\frac{\kappa
r r'}{|r-r'|}\,K_1(\kappa|r-r'|)\right]-\\
\\
-\frac{\sin(F\pi)}{\pi}\int\limits_0^\infty
dy\,\exp\left(-\frac{\kappa^2r r'}{2y}-\frac{r^2+{r'}^2}{2r r'
}\,y\right)\times\\ \times
\left\{\frac{\omega}{2F(1-F)}\left(1+\frac{1}{4y}\right)e^y\int\limits_y^\infty
du\,e^{-u}[(1-F)K_F(u)+F K_{1-F}(u)]-\right.\\ \left.
-\vphantom{\int\limits_y^\infty}
\left[\left(\frac14+y\right)\frac{\left(F-\frac12\right)\omega}{F(1-F)}+m\right][K_F(y)-K_{1-F}(y)]-\omega[(1-F)K_F(y)+F
K_{1-F}(y)]\right\}+\\
+\frac{2\left(F-\frac12\right)^2\sin(F\pi)}{\pi(\tan\nu_\omega+e^{i
F\pi})}\left[(\omega+m)\tan\nu_\omega K_F(\kappa r)K_F(\kappa
r')+(\omega-m)e^{i F\pi}K_{1-F}(\kappa r)K_{1-F}(\kappa
r')\right],
\end{multline}
\begin{multline}\label{c22}
\int\limits_0^{2\pi} d\varphi\,tr\left[\Omega J\,
G^\omega(r,\varphi;r',\varphi)\right]=\frac{e^2}{4\pi}\kappa
r\frac{r+r'}{|r-r'|}\,K_1(\kappa|r-r'|)-
\\-\frac{e^2\sin(F \pi)}{8F(1-F)\pi^2}\int\limits_0^\infty\!\!
dy\,\left(1+\frac{\kappa^2r r'}{4y^2}\!+\!\frac{r^2-{r'}^2}{4r r'
}\right)\exp\left(\!\!-\frac{\kappa^2r
r'}{2y}\!-\!\frac{r^2+{r'}^2}{2r r' }\,y\!\right)\times \\ \times
\left\{e^y\int\limits_y^\infty du\,e^{-u}[(1-F)K_F(u)+F
K_{1-F}(u)]-(2F-1)y[K_F(y)-K_{1-F}(y)]\right\}+\\ +
\frac{e^2\sin(F \pi)}{4\pi^2}\int\limits_0^\infty
dy\,\exp\left(-\frac{\kappa^2r r'}{2y}-\frac{r^2+{r'}^2}{2r
r'}\,y\right)[(1-F)K_F(y)+F K_{1-F}(y)]-\\-
\frac{e^2\left(F-\frac12\right)\sin(F\pi)}{2\pi^2(\tan\nu_\omega+e^{i
F\pi})}\,\kappa r\left[\tan\nu_\omega K_{1-F}(\kappa r)K_F(\kappa
r')-e^{i F\pi}K_F(\kappa r)K_{1-F}(\kappa r')\right],
\end{multline}
where $\kappa=-i k$, $K_\rho(u)$ is the Mackdonald function of
order $\rho$, and $\tan\nu_\omega$ is given by Eq.\eqref{ap13}.
The first terms in the right hand sides of Eqs.\eqref{c18},
\eqref{c20}-\eqref{c22} diverge in the limit $r'\rightarrow r$.
These terms coincide with expressions corresponding to the case of
 absence of the vortex defect (see also Appendix B):
\begin{equation}\label{c23}
\int\limits_0^{2\pi} d\varphi\,tr\left[\Sigma\,
G^\omega(r,\varphi;r',\varphi)\right]\left|_{e\Phi=0}\right.=smK_0(\kappa|r-r'|)\,,
\end{equation}
\begin{equation}\label{c24}
\int\limits_0^{2\pi} d\varphi\,tr\left[\Sigma J\,
G^\omega(r,\varphi;r',\varphi)\right]\left|_{e\Phi=0}\right.=\frac\omega2
K_0(\kappa|r-r'|)\,,
\end{equation}
\begin{equation}\label{c25}
\int\limits_0^{2\pi} d\varphi\,tr\left[J^2\,
G^\omega(r,\varphi;r',\varphi)\right]\left|_{e\Phi=0}\right.=\frac\omega2\left[K_0(\kappa|r-r'|)+4\frac{\kappa
r r'}{|r-r'|}\,K_1(\kappa|r-r'|)\right]\,,
\end{equation}
\begin{equation}\label{c26}
\int\limits_0^{2\pi} d\varphi\,tr\left[\Omega J\,
G^\omega(r,\varphi;r',\varphi)\right]\left|_{e\Phi=0}\right.=\frac{e^2}{4\pi}\kappa
r\frac{r+r'}{|r-r'|} K_1(\kappa|r-r'|)\,,
\end{equation}
and, otherwise,
\begin{equation}\label{c27}
\int\limits_0^{2\pi} d\varphi\,tr\left[\Lambda\,
G^\omega(r,\varphi;r',\varphi)\right]\left|_{e\Phi=0}\right.=\int\limits_0^\infty
d\varphi\,tr\left[\Omega\,
G^\omega(r,\varphi;r',\varphi)\right]\left|_{e\Phi=0}\right.=0\,.
\end{equation}
Thus, quantities \eqref{c17} - \eqref{c22} are made finite in the
limit $r'\rightarrow r$ by subtracting expressions corresponding
to the case of absence of the vortex defect. Integrating over
radial variables, we get the renormalized traces:
\begin{multline}\label{c28}
 Tr\,\Lambda(H-\omega)^{-1}\equiv\int\limits_0^\infty dr\,r\int\limits_0^{2\pi} d\varphi\,tr\left[\Lambda\,
G^\omega(r,\varphi;r,\varphi)\right]=\\=
\frac{s}{\omega^2-m^2}\left[\frac{F^2(\omega+m)\tan\nu_\omega-(1-F)^2(\omega-m)e^{i
F\pi}}{\tan\nu_\omega+e^{i
F\pi}}-\frac23\left(F-\frac12\right)F(1-F)\omega\right]\,,
\end{multline}
\begin{multline}\label{c29}
 Tr\,\Sigma(H-\omega)^{-1}\equiv\int\limits_0^\infty dr\,r\int\limits_0^{2\pi} d\varphi\,\left\{tr\left[\Sigma\,
G^\omega(r,\varphi;r,\varphi)\right]-tr\left[\Sigma\,
G^\omega(r,\varphi;r,\varphi)\right]\left|_{e\Phi=0}\right.\right\}=\\
=-\frac12\,\frac{s}{\omega^2-m^2}\left[\frac{F(\omega+m)\tan\nu_\omega-(1-F)(\omega-m)e^{i
F\pi}}{\tan\nu_\omega+e^{i F\pi}}-F(1-F)m\right]\,,
\end{multline}
\begin{multline}\label{c30}
 Tr\,\Omega(H-\omega)^{-1}\equiv\int\limits_0^\infty dr\,r\int\limits_0^{2\pi} d\varphi\,tr\left[\Omega\,
G^\omega(r,\varphi;r,\varphi)\right]=\\=
-\frac{e^2}{6\pi}\,\frac{s
F(1-F)}{\omega^2-m^2}\,\frac{(1+F)\tan\nu_\omega-(2-F)e^{i
F\pi}}{\tan\nu_\omega+e^{i F\pi}}\,,
\end{multline}
\begin{multline}\label{c31}
 Tr\,\Sigma J(H-\omega)^{-1}\equiv\int\limits_0^\infty dr\,r\int\limits_0^{2\pi} d\varphi\,\left\{tr\left[\Sigma J\,
G^\omega(r,\varphi;r,\varphi)\right]-tr\left[\Sigma J\,
G^\omega(r,\varphi;r,\varphi)\right]\left|_{e\Phi=0}\right.\right\}=\\
=\frac12\,\frac{F-\frac12}{\omega^2-m^2}\frac{F(\omega+m)\tan\nu_\omega-(1-F)(\omega-m)e^{i
F\pi}}{\tan\nu_\omega+e^{i
F\pi}}+\frac14\,\frac{F(1-F)}{\omega^2-m^2}\left[\omega-\frac43(F-\frac12)m\right]\,,
\end{multline}
\begin{multline}\label{c32}
 Tr\,J^2(H-\omega)^{-1}\equiv\int\limits_0^\infty dr\,r\int\limits_0^{2\pi} d\varphi\,\left\{tr\left[ J^2\,
G^\omega(r,\varphi;r,\varphi)\right]-tr\left[J^2\,
G^\omega(r,\varphi;r,\varphi)\right]\left|_{e\Phi=0}\right.\right\}=\\
=-\,\frac{\left(F-\frac12\right)^2}{\omega^2-m^2}\frac{F(\omega+m)\tan\nu_\omega+(1-F)(\omega-m)e^{i
F\pi}}{\tan\nu_\omega+e^{i
F\pi}}+\\+\frac12\frac{F(1-F)}{\omega^2-m^2}\left\{\left[\frac12-F(1-F)\right]\omega-\frac43(F-\frac12)m\right\}\,,
\end{multline}
\begin{multline}\label{c33}
 Tr\,\Omega J(H-\omega)^{-1}\equiv\int\limits_0^\infty dr\,r\int\limits_0^{2\pi} d\varphi\,\left\{tr\left[\Omega J\,
G^\omega(r,\varphi;r,\varphi)\right]-tr\left[\Omega J\,
G^\omega(r,\varphi;r,\varphi)\right]\left|_{e\Phi=0}\right.\right\}=\\=
\frac{e^2}{8\pi}\,\frac{
F(1-F)}{\omega^2-m^2}\,\frac{F(1+F)\tan\nu_\omega+(1-F)(2-F)e^{i
F\pi}}{\tan\nu_\omega+e^{i F\pi}}\,,
\end{multline}
where the integration is performed at $Re \,\kappa>\left|Im\,
\kappa\right|$ and, then, is continued analytically to half-plane
$Re\, \kappa>0$ $(Im \,k>0)$ which corresponds to the whole plane
of complex $\omega$. For completeness, we present here the result
of Ref.\cite{SiG}:
\begin{multline}\label{c34}
 Tr\,(H-\omega)^{-1}\equiv\int\limits_0^\infty dr\,r\int\limits_0^{2\pi}
 d\varphi\,\left[tr\,G^\omega(r,\varphi;r,\varphi)-tr\,
G^\omega(r,\varphi;r,\varphi)\left|_{e\Phi=0}\right.\right]=\\
=-\frac{1}{\omega^2-m^2}\left[\frac{F(\omega+m)\tan\nu_\omega+(1-F)(\omega-m)e^{i
F\pi}}{\tan\nu_\omega+e^{i F\pi}}-F(1-F)\omega\right]\,.
\end{multline}

There are remarkable relations among different traces. In
particular, summing Eqs.\eqref{c28} and \eqref{c29} we get
\begin{equation}\label{c35}
 Tr\,J(H-\omega)^{-1}=-s\left(F-\frac12\right)Tr\,(H-\omega)^{-1}+\frac{s
 F(1-F)}{\omega^2-m^2}\left[\frac13\left(F-\frac12\right)\omega+\frac12m\right]\,.
\end{equation}
Trace \eqref{c30} is expressed through traces \eqref{c29} and
\eqref{c34}:
\begin{multline}\label{c36}
 \frac{2\pi m}{e^2}\,Tr\,\Omega(H-\omega)^{-1}=\frac
 s4\,Tr\,(H-\omega)^{-1}-\left(F-\frac12\right)Tr\,\Sigma(H-\omega)^{-1}+\\
 +\frac14\,\frac{s
 F(1-F)}{\omega^2-m^2}\left[\omega+\frac23\left(F-\frac12\right)m\right]\,.
\end{multline}
We list also some other relations:
\begin{equation}\label{c37}
Tr\,J^2(H-\omega)^{-1}=\left(F-\frac12\right)^2Tr\,(H-\omega)^{-1}+\frac12\,\frac{
 F(1-F)}{\omega^2-m^2}\left[F(1-F)\omega-\frac43\left(F-\frac12\right)m\right]\,,
\end{equation}
\begin{multline}\label{c38}
Tr\,\Lambda
J(H-\omega)^{-1}=-s\left(F-\frac12\right)\,Tr\,\Lambda(H-\omega)^{-1}-\\-\frac13\,\frac{
 F(1-F)}{\omega^2-m^2}\left\{\frac12[1-F(1-F)]\omega+\left(F-\frac12\right)m\right\}\,,
\end{multline}
\begin{equation}\label{c39}
Tr\,\Sigma
J(H-\omega)^{-1}=-s\left(F-\frac12\right)\,Tr\,\Sigma(H-\omega)^{-1}+\frac14\,\frac{
 F(1-F)}{\omega^2-m^2}\left[\omega+\frac23\left(F-\frac12\right)m\right]\,,
\end{equation}
\begin{equation}\label{c40}
Tr\,\Omega
J(H-\omega)^{-1}=-s\left(F-\frac12\right)Tr\,\Omega(H-\omega)^{-1}+\frac{e^2}{12\pi}\,\frac{
F(1-F)}{\omega^2-m^2}\,\left[1+\frac12F(1-F)\right].
\end{equation}


\section{Averages}
Similar to Eq.\eqref{c1}, we get the thermal averages of orbital
angular momentum
\begin{equation}\label{d1}
L(T)=-\frac12\int\limits_C\frac{d\omega}{2\pi
i}\tanh\left(\frac12\beta\omega\right)\,Tr\,\Lambda(H-\omega)^{-1}\,,
\end{equation}
and spin
\begin{equation}\label{d2}
S(T)=-\frac12\int\limits_C\frac{d\omega}{2\pi
i}\tanh\left(\frac12\beta\omega\right)\,Tr\,\Sigma(H-\omega)^{-1}\,,
\end{equation}
where $C$ is the contour $(-\infty+i0,+\infty+i0)$ and
$(+\infty-i0,-\infty-i0)$ in the complex $\omega$-plane. Using
Eqs.\eqref{c28} and \eqref{c29} and deforming the contour around
the cuts and poles on the real axis, we obtain the following
expressions for the averages as real integrals:
\begin{multline}\label{d3}
L(T)=s\frac{\sin(F\pi)}{\pi}\int\limits_0^\infty\frac{du}{u\sqrt{u+1}}\tanh\left(\frac12\beta
m\sqrt{u+1}\right)\times\\ \times
\frac{F^2u^FA+(1-F)^2u^{1-F}A^{-1}+u\left\{\left[\frac12-F(1-F)\right](u^FA+u^{1-F}A^{-1})-(2F-1)\cos(F\pi)\right\}}
{[u^FA-u^{1-F}A^{-1}+2\cos(F\pi)]^2+4(u+1)\sin^2(F\pi)}+\\ +\frac
s4\,[1-{\rm{sgn}}(A)]\frac{[1-2F(1-F)]E_{BS}+(2F-1)m}{(2F-1)E_{BS}+m}\tanh\left(\frac12\beta
E_{BS}\right)\,,
\end{multline}
and
\begin{multline}\label{d4}
S(T)=-s\frac{\sin(F\pi)}{2\pi}\int\limits_0^\infty\frac{du}{u\sqrt{u+1}}\tanh\left(\frac12\beta
m\sqrt{u+1}\right)\times\\\times
\frac{Fu^FA+(1-F)u^{1-F}A^{-1}+u\left[\frac12u^FA+\frac12u^{1-F}A^{-1}-(2F-1)\cos(F\pi)\right]}
{[u^FA-u^{1-F}A^{-1}+2\cos(F\pi)]^2+4(u+1)\sin^2(F\pi)}-\\ -\frac
s8\,[1-{\rm{sgn}}(A)]\frac{E_{BS}+(2F-1)m}{(2F-1)E_{BS}+m}\tanh\left(\frac12\beta
E_{BS}\right)+\frac s4F(1-F)\tanh\left(\frac12\beta m\right)\,,
\end{multline}
where
\begin{equation}\label{d5}
A=2^{1-2F}\frac{\Gamma(1-F)}{\Gamma(F)}\tan\left(s\frac\Theta2+\frac\pi4\right)\,,
\end{equation}
$\Gamma(u)$ is the Euler gamma function, $E_{BS}$ is the energy of
the bound state in the one-particle spectrum, which is determined
as a real root of algebraic equation (for details see
Ref.\cite{Si7})
\begin{equation}\label{d6}
\frac{(1-m^{-1}E_{BS})^F}{(1+m^{-1}E_{BS})^{1-F}}\,A=-1\,;
\end{equation}
note that the bound state exists at $\cos \Theta<0$ $(A<0)$, and
its energy is zero at $A=-1$, and, otherwise, one has
$0<|E_{BS}|<|m|$ and
\begin{equation}\label{d7}
{\rm{sgn}}(E_{BS})=\frac12\,{\rm{sgn}}(m)[\,{\rm{sgn}}(1+A^{-1})-{\rm{sgn}}(1+A)]\,.
\end{equation}

Summing Eqs.\eqref{d3} and \eqref{d4} we get the thermal average
of total angular momentum, which can be written in the form
(compare with Eq.\eqref{c35}):
\begin{equation}\label{d8}
M(T)=-s\left(F-\frac12\right)N(T)+\frac s4
F(1-F)\tanh\left(\frac12\beta m\right)\,,
\end{equation}
where the thermal average of fermion number (i. e. electric charge
divided by $e$) is given by expression (see Ref.\cite{SiG}):
\begin{multline}\label{d9}
N(T)=-\frac{\sin(F\pi)}{\pi}\int\limits_0^\infty\frac{du}{u\sqrt{u+1}}\tanh\left(\frac12\beta
m\sqrt{u+1}\right)\times\\\times
\frac{Fu^FA-(1-F)u^{1-F}A^{-1}+u\left[\left(F-\frac12\right)(u^FA+u^{1-F}A^{-1})-\cos(F\pi)\right]}
{[u^FA-u^{1-F}A^{-1}+2\cos(F\pi)]^2+4(u+1)\sin^2(F\pi)}\,-\\-
\frac14\,[1-{\rm{sgn}}(A)]\tanh\left(\frac12\beta E_{BS}\right).
\end{multline}
Note that both $L(T)$ and $S(T)$ are infinite
$\left(\vphantom{\int\limits^\infty}\right.\!$divergent as
integral
$\left.\displaystyle{\int\limits^\infty\frac{du}{u}}\right)$ at
half-integer values of $e\Phi$, unless $A=0$ or $A^{-1}=0$.
However, this divergence cancels in the sum, and one gets
\begin{equation}\label{d10}
\left.M(T)\right|_{F=\frac12}=\frac s{16}\tanh\left(\frac12\beta m
\right).
\end{equation}
In the cases of $A=0$ and $A^{-1}=0$ expressions \eqref{d3} and
\eqref{d4} simplify
\begin{equation}\label{d11}
L(T)=\frac
s2\left(F-\frac12\pm\frac12\right)^2\tanh\left(\frac12\beta m
\right),\quad\Theta=\pm s\frac\pi2\,({\rm{mod}}\,2\pi)\,,
\end{equation}
and
\begin{equation}\label{d12}
S(T)=-\frac
s4\left(F-\frac12\pm\frac12\right)^2\tanh\left(\frac12\beta m
\right),\quad\Theta=\pm s\frac\pi2\,({\rm{mod}}\,2\pi)\,.
\end{equation}

In the limit $T\rightarrow0$ $(\beta\rightarrow\infty)$ we get the
results of Ref.\cite{Si9}:
\begin{equation}\label{d13}
L(0)=\frac{s\,{\rm{sgn}}(m)}{2\pi}\int\limits_1^\infty\frac{d\upsilon}{\sqrt{\upsilon-1}}\,\frac{F^2\upsilon^{-1+F}A+1-2F(1-F)+(1-F)^2\upsilon^{-F}A^{-1}}
{\upsilon^FA+2+\upsilon^{1-F}A^{-1}}\,,
\end{equation}
and
\begin{equation}\label{d14}
S(0)=\frac14s\,{\rm{sgn}}(m)\left[F(1-F)-\frac1\pi
\int\limits_1^\infty\frac{d\upsilon}{\sqrt{\upsilon-1}}\,\frac{F\upsilon^{-1+F}A+1+(1-F)\upsilon^{-F}A^{-1}}
{\upsilon^FA+2+\upsilon^{1-F}A^{-1}}\right].
\end{equation}
In the high-temperature limit the averages tend to zero
\begin{multline}\label{d15}
L(T\rightarrow\infty)=\\=\left\{\begin{array}{lr} {\displaystyle
s\,{\rm sgn}(m)
\frac{\sin(F\pi)}{2\pi}\frac{\Gamma(1-F)}{\Gamma(1+F)}\frac{1-2F(1-F)}{1-2F}
\tan\left(s\frac\Theta2+\frac\pi4\right)\left(\frac{|m|}{k_{B}T}\right)^{1-2F}},
&{\displaystyle0<F<\frac12}\\ {\displaystyle s\,{\rm sgn}(m)
\frac{\sin(F\pi)}{2\pi}\frac{\Gamma(F)}{\Gamma(2-F)}\frac{1-2F(1-F)}{2F-1}
\cot\left(s\frac\Theta2+\frac\pi4\right)\left(\frac{|m|}{k_{B}T}\right)^{2F-1}}
,&{\displaystyle\frac12<F<1}
\end{array}\right.
\end{multline}
and
\begin{multline}\label{d16}
S(T\rightarrow\infty)=\\=\left\{\begin{array}{lr} {\displaystyle
-s\,{\rm sgn}(m)
\frac{\sin(F\pi)}{4\pi}\frac{\Gamma(1-F)}{\Gamma(1+F)}\frac{1}{1-2F}
\tan\left(s\frac\Theta2+\frac\pi4\right)\left(\frac{|m|}{k_{B}T}\right)^{1-2F}},
&{\displaystyle0<F<\frac12}\\ {\displaystyle -s\,{\rm sgn}(m)
\frac{\sin(F\pi)}{4\pi}\frac{\Gamma(F)}{\Gamma(2-F)}\frac{1}{2F-1}
\cot\left(s\frac\Theta2+\frac\pi4\right)\left(\frac{|m|}{k_{B}T}\right)^{2F-1}}
,&{\displaystyle\frac12<F<1}
\end{array}\right.
\end{multline}

In conclusion of this section, let us consider the thermal average
of induced flux \mbox{(times $e$)}
\begin{equation}\label{d17}
O(T)=-\frac12\int\limits_C\frac{d\omega}{2\pi
i}\tanh(\beta\omega)\,Tr\,\Omega(H-\omega)^{-1}\,,
\end{equation}
where $\Omega$ is given by Eq.\eqref{b19}. Using trace identity
\eqref{c36}, we get
\begin{equation}\label{d18}
O(T)=\frac{e^2}{2\pi m}\left[\frac
s4\,N(T)-\left(F-\frac12\right)S(T)+\frac{s}{12}\left(F-\frac12\right)F(1-F)
\tanh\left(\frac12\beta m\right)\right],
\end{equation}
or
\begin{multline}\label{d19}
O(T)=-\frac{e^2}{4\pi m}\,s F(1-F)\left\{\frac{\sin(F\pi)}{\pi}
\int\limits_0^\infty\frac{du}{u\sqrt{u+1}}\tanh\left(\frac12\beta
m\sqrt{u+1}\right)\right.\times \\ \times
\frac{u^FA-u^{1-F}A^{-1}-2u\cos(F\pi)}
{[u^FA-u^{1-F}A^{-1}+2\cos(F\pi)]^2+4(u+1)\sin^2(F\pi)}+\\+\frac12\,[1-{\rm{sgn}}(A)]\left.
\frac{m}{(2F-1)E_{BS}+m}\tanh\left(\frac12\beta
E_{BS}\right)+\frac13\left(F-\frac12\right)\tanh\left(\frac12\beta
m\right)\right\}.
\end{multline}
Unlike $L(T)$ and $S(T)$, $O(T)$ is finite at half-integer values
of $e\Phi$:
\begin{multline}\label{d22}
\left.O(T)\right|_{F=\frac12}=-\frac {e^2}{32\pi m}\left\{[1-{\rm
sgn}(\cos\Theta)]
 \tanh\left(\frac12\beta m\sin\Theta\right)+\right.\\
\left.+\frac{\sin2\Theta}{2\pi}\int\limits_1^\infty\frac{dv}{\sqrt{v(v-1)}}
\frac{\tanh\left({\displaystyle\frac12\beta
m\sqrt{v}}\right)}{v-\sin^2{\Theta}}\right\},
\end{multline}
In the cases of $A=0$ and $A^{-1}=0$, Eq.\eqref{d19} takes form
\begin{equation}\label{d23}
O(T)=-\frac {e^2}{12\pi m}s F
(1-F)\left(F-\frac12\pm\frac32\right)\tanh\left(\frac12\beta m
\right),\quad\Theta=\pm s\frac\pi2\,({\rm{mod}}\,2\pi)\,.
\end{equation}
Note that relation \eqref{d18} at zero temperature was first
obtained in Ref.\cite{Si9}, and expression \eqref{d19} at zero
temperature takes form (see Ref.\cite{Si7})
\begin{equation}\label{d20}
O(0)=-\frac{e^2}{4\pi|m|}\,s
F(1-F)\left[\frac13\left(F-\frac12\right)+\frac1{2\pi}\int\limits_1^\infty
\frac{d\upsilon}{\upsilon\sqrt{\upsilon-1}}\,\frac{\upsilon^{F}A-\upsilon^{1-F}A^{-1}}
{\upsilon^FA+2+\upsilon^{1-F}A^{-1}}\right].
\end{equation}
In the high-temperature limit we get
\begin{multline}\label{d21}
O(T\rightarrow\infty)=-\frac{e^2}{8\pi k_B T}\,s
F(1-F)\left\{\frac12\,[1-{\rm{sgn}}(A)]
\frac{E_{BS}}{(2F-1)E_{BS}+m}+\frac13\left(F-\frac12\right)+\right.\\
\left.+\frac{\sin(F\pi)}{\pi} \int\limits_0^\infty\frac{du}{u}
\frac{u^FA-u^{1-F}A^{-1}-2u\cos(F\pi)}
{[u^FA-u^{1-F}A^{-1}+2\cos(F\pi)]^2+4(u+1)\sin^2(F\pi)}\right\}.
\end{multline}

\section{Correlations}
Similar to Eq.\eqref{c2}, we get the thermal correlations of
fermion number with orbital angular momentum
\begin{equation}\label{e1}
\Delta(T;\hat L,\hat N)=\frac14\int\limits_C\frac{d\omega}{2\pi
i}\,{\rm
sech}^2\left(\frac12\beta\omega\right)Tr\,\Lambda(H-\omega)^{-1},
\end{equation}
and fermion number with spin
\begin{equation}\label{e2}
\Delta(T;\hat S,\hat N)=\frac14\int\limits_C\frac{d\omega}{2\pi
i}\,{\rm
sech}^2\left(\frac12\beta\omega\right)Tr\,\Sigma(H-\omega)^{-1},
\end{equation}
where contour $C$ is defined as above. Using Eqs.\eqref{c28} and
\eqref{c29} and deforming the contour around the cuts and poles on
the real axis, we obtain the following expressions for the
correlations as real integrals:
\begin{multline}\label{e3}
\Delta(T;\hat L,\hat
N)=-\frac{s\,\sin(F\pi)}{2\pi}\int\limits_0^\infty\frac{du}{u}\,{\rm
sech}^2\left(\frac12\beta m\sqrt{u+1}\right)\times\\ \times
\frac{F^2u^FA-(1-F)^2u^{1-F}A^{-1}-u[1-2F(1-F)]\cos(F\pi)}
{[u^FA-u^{1-F}A^{-1}+2\cos(F\pi)]^2+4(u+1)\sin^2(F\pi)}-\\ -\frac
s8\,[1-{\rm{sgn}}(A)]\frac{[1-2F(1-F)]E_{BS}+(2F-1)m}{(2F-1)E_{BS}+m}\,{\rm
sech}^2\left(\frac12\beta E_{BS}\right)+\\+\frac
s6\left(F-\frac12\right)F(1-F){\rm sech}^2\left(\frac12\beta m
\right),
\end{multline}
and
\begin{multline}\label{e4}
\Delta(T;\hat S,\hat
N)=\frac{s\,\sin(F\pi)}{4\pi}\int\limits_0^\infty\frac{du}{u}\,{\rm
sech}^2\left(\frac12\beta m\sqrt{u+1}\right)\times\\ \times
\frac{Fu^FA-(1-F)u^{1-F}A^{-1}-u\cos(F\pi)}
{[u^FA-u^{1-F}A^{-1}+2\cos(F\pi)]^2+4(u+1)\sin^2(F\pi)}+\\ +\frac
s{16}\,[1-{\rm{sgn}}(A)]\frac{E_{BS}+(2F-1)m}{(2F-1)E_{BS}+m}\,{\rm
sech}^2\left(\frac12\beta E_{BS}\right),
\end{multline}
where $A$ is defined by Eq.\eqref{d5}, and $E_{BS}$ is determined
as a real root of Eq.\eqref{d6}.

It should be noted that, if one takes operator
\begin{equation}\label{e5}
\Lambda'=\Lambda+e\Phi=-i\,\textbf{x}\times\mbox{\boldmath
$\partial$},
\end{equation}
and defines the corresponding operator in the second-quantized
theory,
\begin{equation}\label{e6}
{\hat L}'=\frac12\int d^2x\,[\Psi^+,\Lambda'\Psi]_-,
\end{equation}
then correlation $\Delta(T;{\hat L}',\hat N)$ is infinite. To see
this, let us consider quantity (compare with Eq.\eqref{c17})
\begin{multline}\label{e7}
\int\limits_0^{2\pi} d\varphi\,tr\left[\Lambda'\,
G^\omega(r,\varphi;r',\varphi)\right]=\frac{2\sin(F\pi)}{\pi(\tan\nu_\omega+e^{i
F\pi})}\times\\ \times\left[n_0(\omega+m)\tan\nu_\omega K_F(\kappa
r)K_F(\kappa r')+ (n_0+s)(\omega-m)e^{i F\pi}K_{1-F}(\kappa
r)K_{1-F}(\kappa r')\right]-\\
-\frac{\sin(F\pi)}{2F(1-F)\pi}\,\omega\int\limits_0^\infty
dy\,\exp\left(-\frac{\kappa^2r r'}{2y}-\frac{r^2+{r'}^2}{2r r'
}\,y\right)\left\{e\Phi\frac{e^y}{y}\int\limits_y^\infty du
\,e^{-u} [(1-F)K_F(u)\right.+\\+ \left.
FK_{1-F}(u)]-\left[(2F-1)\left(n_0+\frac12\,s\right)+\frac12\,s\right][K_F(y)-K_{1-F}(y)]\right\}+2\omega
e\Phi K_0(\kappa|r-r'|).
\end{multline}
The last term in Eq.\eqref{e7} diverges in the limit $r'\rightarrow
r$, and this divergence can not be compensated by subtraction (as is
the case for Eqs.\eqref{c18},\eqref{c20}-\eqref{c22}), because
Eq.\eqref{c27} holds. Since this divergence is proportional to
$\omega$, it does not contribute to average $L'(T)$, but does
contribute to correlation $\Delta(T;{\hat L}',\hat N)$ yielding a
term which is $-\frac12\,e\Phi\,{\rm sech}^2\left(\frac12\beta
m\right)$ times infinity. Thus, if one accepts finiteness of
correlations as physically plausible condition, then one has to
favour gauge-invariant definition of orbital angular momentum, i.e.
to choose $\hat \Lambda$\eqref{c17} instead of ${\hat
\Lambda}'$\eqref{e5}.

Contrary to the case of averages $L(T)$ and $S(T)$, correlations
$\Delta(T;\hat L,\hat N)$ and $\Delta(T;\hat S,\hat N)$ are finite
at half-integer values of $e\Phi$:
\begin{multline}\label{e12}
 \left.\Delta(T;\hat L,\hat N)\right|_{F=\frac12}=- \left.\Delta(T;\hat S,\hat N)\right|_{F=\frac12}=
 -\frac{\sin(2\Theta)}{32\pi}\int\limits_1^\infty\frac{d\upsilon}{\sqrt{\upsilon-1}}\frac{{\rm
sech}^2\left(\frac12\beta
m\sqrt{\upsilon}\right)}{\upsilon-\sin^2\Theta}-\\-\frac{\sin\Theta}{16}[1-{\rm
sgn}(\cos\Theta)]\,{\rm sech}^2\left(\frac12\beta
m\sin\Theta\right).
\end{multline}
In the case of $A=0$ and $A^{-1}=0$ expressions \eqref{e3} and
\eqref{e4} simplify:
\begin{equation}\label{e14}
 \Delta(T;\hat L,\hat N)=\frac s6 \left(F-\frac12\pm\frac12\right)
 \!\left[1+2\left(F-\frac12\pm\frac12\right)^2\right]{\rm sech}^2\left(\frac12\beta
m\right),\,\,\,\Theta=\pm s\frac\pi2\,({\rm{mod}}\,2\pi),
\end{equation}
and
\begin{equation}\label{e15}
\Delta(T;\hat S,\hat N)=-\frac s4 \left(F-\frac12\pm\frac12\right)
\,{\rm sech}^2\left(\frac12\beta m\right),\quad\Theta=\pm
s\frac\pi2\,({\rm{mod}}\,2\pi).
\end{equation}
In the limit $T\rightarrow0$ $(\beta\rightarrow\infty)$
correlations \eqref{e3} and \eqref{e4} tend exponentially to zero
for almost all values of $\Theta$ with the exception of one
corresponding to the zero bound state energy, $E_{BS}=0$ $(A=-1)$:
\begin{equation}\label{e8}
\Delta(0;\hat L,\hat N)=-\frac12\Delta(0;\hat S,\hat
N)=\left\{\begin{array}{lr} 0, &A\neq-1\\ {\displaystyle -\frac
s2\left(F-\frac12\right)} ,&A=-1
\end{array}\right..
\end{equation}
In the high-temperature limit the correlations tend to finite
values:
\begin{multline}\label{e10}
 \Delta(\infty;\hat L,\hat
 N)=-\frac
s8\,[1-{\rm{sgn}}(A)]\frac{[1-2F(1-F)]E_{BS}+(2F-1)m}{(2F-1)E_{BS}+m}+\frac
s6\left(F-\frac12\right)F(1-F)-\\-\frac{s\,\sin(F\pi)}{2\pi}\int\limits_0^\infty\frac{du}{u}
\frac{F^2u^FA-(1-F)^2u^{1-F}A^{-1}-u[1-2F(1-F)]\cos(F\pi)}
{[u^FA-u^{1-F}A^{-1}+2\cos(F\pi)]^2+4(u+1)\sin^2(F\pi)},
\end{multline}
and
\begin{multline}\label{e11}
\Delta(\infty;\hat S,\hat N)=\frac
s{16}\,[1-{\rm{sgn}}(A)]\frac{E_{BS}+(2F-1)m}{(2F-1)E_{BS}+m}+\\+
\frac{s\,\sin(F\pi)}{4\pi}\int\limits_0^\infty\frac{du}{u}
\frac{Fu^FA-(1-F)u^{1-F}A^{-1}-u\cos(F\pi)}
{[u^FA-u^{1-F}A^{-1}+2\cos(F\pi)]^2+4(u+1)\sin^2(F\pi)}\,.
\end{multline}

Summing Eqs.\eqref{e3} and \eqref{e4}, we get the thermal
correlation of two conserved observables, total angular momentum
and fermion number, which can be recast in the form:
\begin{equation}\label{e16}
\Delta(T;\hat M,\hat N)=-s\left(F-\frac12\right)\Delta(T;\hat
N,\hat N)-\frac s{12}\left(F-\frac12\right)F(1-F){\rm
sech}^2\left(\frac12\beta m \right),
\end{equation}
where
\begin{multline}\label{e17}
\Delta(T;\hat N,\hat
N)=\frac{\sin(F\pi)}{2\pi}\int\limits_0^\infty\frac{du}{u}\,{\rm
sech}^2\left(\frac12\beta m\sqrt{u+1}\right)\times\\ \times
\frac{Fu^FA+(1-F)u^{1-F}A^{-1}-u(2F-1)\cos(F\pi)}
{[u^FA-u^{1-F}A^{-1}+2\cos(F\pi)]^2+4(u+1)\sin^2(F\pi)}+\\ +\frac
1{8}\,[1-{\rm{sgn}}(A)]\,{\rm sech}^2\left(\frac12\beta
E_{BS}\right)-\frac14 F(1-F)\,{\rm sech}^2\left(\frac12\beta
m\right)
\end{multline}
is the quadratic fluctuation of fermion number which was first
computed in Ref.\cite{SiG}. Note that correlation $\Delta(T;\hat
M,\hat N)$ vanishes at half-integer values of $e\Phi$.

Using trace identities \eqref{c38} and \eqref{c39}, we get the
thermal correlations of total angular momentum with orbital
angular momentum
\begin{equation}\label{e18}
\Delta(T;\hat L,\hat M)=-s\left(F-\frac12\right)\Delta(T;\hat
L,\hat N)+\frac1{24}[1-F(1-F)]F(1-F)\,{\rm
sech}^2\left(\frac12\beta m \right),
\end{equation}
and total angular momentum with spin
\begin{equation}\label{e19}
\Delta(T;\hat S,\hat M)=-s\left(F-\frac12\right)\Delta(T;\hat
S,\hat N)-\frac1{16}F(1-F)\,{\rm sech}^2\left(\frac12\beta m
\right).
\end{equation}

Using trace identity \eqref{c36}, we get the thermal correlation
of fermion number with induced flux multiplied by $e$
\begin{equation}\label{e20}
\Delta(T;\hat O,\hat N)=\frac{e^2}{2\pi m }\left[\frac s4
\Delta(T;\hat N,\hat N)-\left(F-\frac12\right)\Delta(T;\hat S,\hat
N)-\frac s{16}F(1-F)\,{\rm sech}^2\left(\frac12\beta m
\right)\right],
\end{equation}
or in the explicit form
\begin{multline}\label{e21}
\Delta(T;\hat O,\hat N)=\frac{e^2\sin(F\pi)}{2(2\pi)^2}\frac{s
F(1-F)}{m}\int\limits_0^\infty\frac{du}{u}\,{\rm
sech}^2\left(\frac12\beta m\sqrt{u+1}\right)\times\\ \times
\frac{u^FA+u^{1-F}A^{-1}}
{[u^FA-u^{1-F}A^{-1}+2\cos(F\pi)]^2+4(u+1)\sin^2(F\pi)}+\\ +\frac
{e^2}{16\pi}\,[1-{\rm{sgn}}(A)]\frac{s
F(1-F)}{(2F-1)E_{BS}+m}\,{\rm sech}^2\left(\!\frac12\beta
E_{BS}\!\right)-\frac{e^2s F(1-F)}{16\pi m}\,{\rm
sech}^2\left(\!\frac12\beta m\!\right).
\end{multline}
At half-integer values of $e\Phi$ we get
\begin{multline}\label{e23}
 \left.\Delta(T;\hat O,\hat
 N)\right|_{F=\frac12}=\frac{s\,e^2\cos\Theta}{4(4\pi)^2m}\int\limits_1^\infty
 \frac{d\upsilon}{\sqrt{\upsilon-1}}\frac{\,{\rm
sech}^2\left(\frac12\beta
m\sqrt{\upsilon}\right)}{\upsilon-\sin^2\Theta}+\\+\frac{s\,e^2}{64
\pi m}\,[1-{\rm sgn}(\cos\Theta)]\,{\rm sech}^2\left(\frac12\beta
m\sin\Theta\right)-\frac{s\,e^2}{64 \pi m}\,{\rm
sech}^2\left(\frac12\beta m\right).
\end{multline}
Correlation $\Delta(T;\hat O,\hat N)$ vanishes in the cases of
$A=0$ and $A^{-1}=0$. In the zero-temperature limit we get
\begin{equation}\label{e24}
\Delta(0;\hat O,\hat N)=\left\{\begin{array}{lr} 0, &A\neq-1\\
{\displaystyle\frac{e^2s F(1-F)}{8\pi m}} ,&A=-1
\end{array}\right..
\end{equation}
In the high-temperature limit correlation \eqref{e21} tends to a
finite value
\begin{multline}\label{e25}
\Delta(\infty;\hat O,\hat N)=\frac
{e^2}{16\pi}\,[1-{\rm{sgn}}(A)]\frac{s
F(1-F)}{(2F-1)E_{BS}+m}-\frac{e^2s F(1-F)}{16\pi
m}+\\+\frac{e^2\sin(F\pi)}{2(2\pi)^2}\frac{s
F(1-F)}{m}\int\limits_0^\infty\frac{du}{u}\,\frac{u^FA+u^{1-F}A^{-1}}
{[u^FA-u^{1-F}A^{-1}+2\cos(F\pi)]^2+4(u+1)\sin^2(F\pi)}.
\end{multline}

Using trace identity \eqref{c40}, we get the thermal correlation
of total angular momentum with induced flux multiplied by $e$
\begin{equation}\label{e26}
\Delta(T;\hat O,\hat M)=-s\left(F-\frac12\right)\Delta(T;\hat
O,\hat N).
\end{equation}
Thus, this correlation vanishes at half-integer values of $e\Phi$.

\section{Nonnegativeness of quadratic fluctuations}
As we have seen in the previous section, a gauge-invariant
definition of angular momentum is required by the finiteness of
the correlation of angular momentum with fermion number. In the
present section we shall consider further restrictions which are
imposed by the nonnegativeness of quadratic fluctuations.

With the use of trace identity \eqref{c37}, the thermal quadratic
fluctuation of total angular momentum, Eq.\eqref{c2}, is expressed
through that of fermion number:
\begin{equation}\label{f1}
\Delta(T;\hat M,\hat M)=\left(F-\frac12\right)^2\Delta(T;\hat
N,\hat N)-\frac18F^2(1-F)^2{\rm sech}^2\left(\frac12\beta
m\right),
\end{equation}
where $\Delta(T;\hat N,\hat N)$ is given by Eq.\eqref{e17}. The
thermal quadratic fluctuation of fermion number was analyzed in
detail in Ref.\cite{SiG}. In particular, in the high-temperature
limit we get
\begin{equation}\label{f2}
\Delta(\infty;\hat N,\hat N) =\left\{
\begin{array}{lr}
\left.\begin{array}{lr}
{\displaystyle\vphantom{\int}\!\!\!\frac{1}4(1-F)^2},&
{\displaystyle\Theta\neq
s\frac\pi2\,({\rm mod}\,2\pi)}\vspace{0.3em}\\
{\displaystyle\vphantom{\int}\!\!\!\frac{1}4F^2},&
{\displaystyle\Theta=
s\frac\pi2\,({\rm mod}\,2\pi)}\vspace{0.3em}\phantom{ss\!}\\
\end{array}\right\},&{\displaystyle  0<F\leq\frac12}\vspace{0.3em}\\
\left.\begin{array}{lr}
{\displaystyle\vphantom{\int}\!\!\!\frac{1}4F^2},&
{\displaystyle\Theta\neq
-s\frac\pi2\,({\rm mod}\,2\pi)}\vspace{0.3em}\\
{\displaystyle\vphantom{\int}\!\!\!\frac{1}4(1-F)^2},&
{\displaystyle\Theta=
-s\frac\pi2\,({\rm mod}\,2\pi)}\vspace{0.3em}\\
\end{array}\right\},&{\displaystyle  \frac12\leq F<1}\\
\end{array}
\right.
\end{equation}
which is obviously positive. Substituting Eq.\eqref{f2} into
Eq.\eqref{f1}, we get
\begin{equation}\label{f3}
\Delta(\infty;\hat M,\hat M) =\left\{
\begin{array}{lr}
\left.\begin{array}{lr}
{\displaystyle\vphantom{\int}\!\!\!\frac{1}8(1-F)^2\left[(1-F)^2-\frac12\right]},&
{\displaystyle\Theta\neq
s\frac\pi2\,({\rm mod}\,2\pi)}\vspace{0.3em}\\
{\displaystyle\vphantom{\int}\!\!\!\frac{1}8F^2\left(F^2-\frac12\right)},&
{\displaystyle\Theta=
s\frac\pi2\,({\rm mod}\,2\pi)}\vspace{0.3em}\phantom{ss\!}\\
\end{array}\right\},&{\displaystyle  0<F\leq\frac12}\vspace{0.3em}\\
\left.\begin{array}{lr}
{\displaystyle\vphantom{\int}\!\!\!\frac{1}8F^2\left(F^2-\frac12\right)},&
{\displaystyle\Theta\neq
-s\frac\pi2\,({\rm mod}\,2\pi)}\vspace{0.3em}\\
{\displaystyle\vphantom{\int}\!\!\!\frac{1}8(1-F)^2\left[(1-F)^2-\frac12\right]},&
{\displaystyle\Theta=
-s\frac\pi2\,({\rm mod}\,2\pi)}\vspace{0.3em}\\
\end{array}\right\},&{\displaystyle  \frac12\leq F<1}\\
\end{array}
\right.
\end{equation}
which is positive at $0<F<1-2^{-1/2}$ and $2^{-1/2}<F<1$
${\displaystyle\left(\Theta\neq \frac\pi2\,({\rm
mod}\,\pi)\right)}$ only, and is nonpositive otherwise.

The cause of negativeness of $\Delta(\infty;\hat M,\hat M)$ is
quite understandable from the mathematical point of view. Although
operator $J^2$, as a square of self-adjoint operator $J$, is
nonnegative definite, appropriate spectral density $\tau_{J^2}(E)$
might be not, and the latter results in negativeness of
$\Delta(T;\hat M,\hat M)$, see Eq.\eqref{intr22}. Nonnegativeness
of $\tau_{J^2}(E)$ is rooted in the procedure of taking a
functional trace of the resolvent kernel, with regularization and
renormalization involved in the case of infinite space: initially,
the trace is positive but divergent (see Eq.\eqref{c21} at
$r'\rightarrow r$), the comparative trace which corresponds to the
case of absence of the defect is also positive but divergent (see
Eq.\eqref{c25} at $r'\rightarrow r$), then the difference of the
above two traces appears to be finite but not positive. Namely the
same is the mechanism of the appearance of the negative vacuum
energy density, which is widely known as the Casimir effect
\cite{Cas}: the vacuum energy density in an infinite space bounded
by two parallel plates is positive but divergent, the vacuum
energy density in an infinite unbounded space is subtracted, and
the result is finite but negative. There is a physically plausible
interpretation of negativeness of the vacuum energy density,
linking it to a force of attraction between two plates. Returning
now to negativeness of the thermal quadratic fluctuation, we have
not found, up to now, any physically plausible interpretation of
this mathematically feasible effect. Thus, we have to stick to the
paradigma that the quadratic fluctuation of the physically
meaningful observable is to be nonnegative.

Let us recall that, in the case of planar rotationally symmetric
system considered in the present paper, we have two conserved
observables which are fermion number and total angular momentum.
Fermion number is defined uniquely: operator $\hat N$ in the
second-quantized theory corresponds to unity operator $I$ in the
first-quantized theory. As to total angular momentum, situation is
different: so far we have chosen operator $\hat M$ corresponding
to $J$\eqref{b10}, but, equally as well, the choice can be a
superposition of $\hat M$ and $\hat N$, $\hat M +\Xi\hat N$, see
Eq.\eqref{b16}, with $\Xi$ being a function of the parameters of
the vortex defect. We shall use this ambiguity and fix $\Xi$ by
the requirement that the quadratic fluctuation of modified
(improved) total angular momentum behave qualitatively in a
similar manner as that of fermion number.

First, it is straightforward to get general relation
\begin{multline}\label{f4}
\Delta(T;\hat M +\Xi\hat N,\hat M +\Xi\hat
N)=\left(F-\frac12-s\Xi\right)^2\Delta(T;\hat N,\hat
N)-\\-\frac12\left[\frac s3
\Xi\left(F-\frac12\right)+\frac14F(1-F)\right]F(1-F)\,{\rm
sech}^2\left(\frac12\beta m\right).
\end{multline}
Then, one should note that, without a loss of generality, all
possible values of $\Xi$ can be restricted to interval
$0\leq|\Xi|\leq1/2$, since shift $\Xi\rightarrow\Xi\pm1$ yields
the same spectrum of the angular momentum operator in the
first-quantized theory. Also, in view of the Bohm-Aharonov effect
\cite{Aha}, we suppose that $\Xi$ depends on fractional part of
$e\Phi$ rather than $e\Phi$ itself, or, in other words, it depends
on $F$\eqref{c10}. Since $F\rightarrow1-F$ as $s\rightarrow-s$,
parameter $\Xi$ has to depend on $s(F-1/2)$. Taking all the above
into account, we fix $\Xi=\Xi_F$, where
\begin{equation}\label{f5}
\Xi_F=\left\{\begin{array}{lr} {\displaystyle-\frac12\,s\,{\rm sgn}\left(F-\frac12\right)},&{\displaystyle F\neq\frac12}\\
{\displaystyle\frac12},&{\displaystyle F=\frac12}
\end{array}\right.,
\end{equation}
and define  the improved total angular momentum operator as
\begin{equation}\label{f6}
    \hat R=\hat M+\Xi_F\hat N.
\end{equation}
The corresponding quadratic fluctuation takes form
\begin{multline}\label{f7}
\Delta(T;\hat R,\hat
R)=\left(\left|F-\frac12\right|+\frac12\right)^2\Delta(T;\hat
N,\hat N)+\\+\frac14\left[\frac 13
\left|F-\frac12\right|-\frac12F(1-F)\right]F(1-F)\,{\rm
sech}^2\left(\frac12\beta m\right).
\end{multline}
The high-temperature limit of Eq.\eqref{f7} is positive, and the
behaviour of $\Delta(T;\hat R,\hat R)$ at finite temperatures is
qualitatively the same as that of $\Delta(T;\hat N,\hat N)$.

\section{Summary and discussion}
In the present paper we continue a study of the properties of an
ideal gas of twodimensional relativistic massive electrons in the
background of a static magnetic vortex defect, which was started
in Ref.\cite{SiG}. We find that this system at thermal equilibrium
acquires, in addition to fermion number considered in
Ref.\cite{SiG}, the following nontrivial characteristics: orbital
angular momentum \eqref{d3}, spin \eqref{d4}, and induced magnetic
flux (times e) \eqref{d19}. The local features of the field
strength in the interior of the vortex are exhibited by
self-adjoint extension parameter $\Theta$ which labels boundary
conditions at the location of the vortex defect, and arbitrary
values of vortex flux $\Phi$ are permitted; our results are
periodic in $\Theta$ with period $2\pi$ at fixed $\Phi$ and
periodic in $\Phi$ with period $e^{-1}$ at fixed $\Theta$. Orbital
angular momentum and spin are odd and induced flux is even under
transition to the inequivalent representation of the Clifford
algebra ($s\rightarrow-s$ or $m\rightarrow-m$). In the
zero-temperature limit the results of Refs.\cite{Si9} and
\cite{Si7} are recovered, see Eqs.\eqref{d13}, \eqref{d14} and
\eqref{d20}, whereas in the high-temperature limit all averages
vanish, see Eqs.\eqref{d15}, \eqref{d16} and \eqref{d21}.

The key point in this study is played by Section 3, where the
appropriately renormalized traces of the resolvent operator are
obtained, see Eqs.\eqref{c28}-\eqref{c33}. Thermal averages are
then computed as integrals over a contour in the complex energy
plane, see Eqs.\eqref{d1}, \eqref{d2} and \eqref{d17}. Moreover,
the knowledge of the resolvent traces allows one to compute also
thermal correlations of conserved and nonconserved observables and
thermal quadratic fluctuations of conserved observables. In
particular, we have computed the correlations of fermion number
with orbital angular momentum, spin and induced flux multiplied by
$e$, see Eqs.\eqref{e3}, \eqref{e4} and \eqref{e21}. These
correlations vanish at zero temperature unless the bound state
energy in the one-particle spectrum vanishes ($A=-1$), see
Eqs.\eqref{e8} and \eqref{e24}. The high-temperature limits of the
correlations are given by Eqs.\eqref{e10}, \eqref{e11} and
\eqref{e25}. Note that correlations \eqref{e3} and \eqref{e4} are
even and correlation \eqref{e21} is odd under transition to the
inequivalent representation of the Clifford algebra
($s\rightarrow-s$ or $m\rightarrow-m$).

It should be emphasized that, owing to the gauge-invariant
definition of orbital angular momentum \eqref{b17}, the
correlation of orbital angular momentum with fermion number is
finite. As it is shown in Section 5, another definition of orbital
angular momentum (see Eq.\eqref{e5}) results in the infinity of
the appropriate correlation, which can be regarded as unphysical.

To illustrate the behaviour of an average and a correlation as
functions of the boundary parameter $\Theta$, we depict these
quantities on Figs.1-3 for one observable, induced flux multiplied
by $e$ (given by operator $\hat O$ \eqref{intr14} with $\Omega$
\eqref{b19}), at several values of the vortex flux. Here
quantities $s|m|e^{-2}\hat O(T)$ and $sme^{-2}\Delta(T;\hat O,
\hat N)$ are along the ordinate axes, and quantity
$s\Theta\pi^{-1}$ is along the abscissa axes. Values
$(k_BT/|m|)=5^{-1}$, 1, 5 correspond to two dashed (with longer
and shorter dashes) and one dotted lines, and values $T=0$ and
$T=\infty$ correspond to solid lines; the latter cannot lead to
confusion, since, as it has been already noted, the average at
$T=\infty$ vanishes everywhere, while the correlation at $T=0$
vanishes almost everywhere with the exception of one point
($A=-1$). Our plots correspond to three values of F \eqref{c10}
from interval $0<F\leq 1/2$, whereas interval $1/2<F<1$ can be
considered by taking in view that $s|m|e^{-2}\hat O(T)\rightarrow-
s|m|e^{-2}\hat O(T)$ and $sme^{-2}\Delta(T;\hat O, \hat
N)\rightarrow sme^{-2}\Delta(T;\hat O, \hat N)$ at
$F\rightarrow1-F$ and $\Theta\rightarrow-\Theta$.

As is seen from Figs.1-3, the average at zero temperature is
characterized by a jump with a cusp at the point corresponding to
the zero bound state energy $(A=-1)$. As temperature increases,
this jump is smoothed out, while extremum evolves close to
$\Theta=s\frac\pi2\,({\rm mod}\,\,2\pi)$ ($A^{-1}=0$) in the case
of $0<F\leq1/2$ and to $\Theta=-s\frac\pi2\,({\rm mod}\,\,2\pi)$
($A=0$) in the case of $1/2\leq F<1$. As temperature departs from
zero, the correlation develops a maximum at $A=-1$, which persists
in the case of $F=1/2$ and is shifted to the left in the case of
$0<F<1/2$ and to the right in the case of $1/2<F<1$, with further
increase of temperature. At non-zero temperature the correlation
becomes negative at $\cos\Theta>0$ $(A>0)$, i.e. in the case of
absence of the bound state in the one-particle spectrum.

Due to  rotational invariance of the system considered, there is
an additional to fermion number conserved observable --- total
angular momentum. It is natural to take in this capacity the sum
of orbital angular momentum and spin: $\hat M=\hat L +\hat S$. The
appropriate average is related to the average fermion number, see
Eqs.\eqref{d8} and \eqref{d9}, and the correlation with fermion
number is related to the quadratic fluctuation of fermion number,
see Eqs.\eqref{e16} and \eqref{e17}. The latter relations are a
consequence of trace identity \eqref{c35}, and, using trace
identities \eqref{c38}-\eqref{c40}, we get the correlations of
total angular momentum with orbital angular momentum, spin and
induced flux times $e$, see Eqs.\eqref{e18},\eqref{e19} and
\eqref{e26}. Using trace identity \eqref{c37}, we get the
quadratic fluctuation of total angular momentum \eqref{f1}.
However, the negativeness of this fluctuation signifies that the
above defined total angular momentum can not be regarded as a
physically meaningful observable.

To remedy the situation, in Section 6 we introduce improved total
angular momentum as a sum of naive total angular momentum and
fermion number with the dependent on the vortex flux coefficient,
see Eqs.\eqref{f5} and \eqref{f6}; the quadratic fluctuation of
the improved observable is given by Eq.\eqref{f7}. On Figs.4-6 we
plot averages and quadratic fluctuations of naive $(\hat M)$ and
improved $(\hat R)$ total angular momenta for several values of
$F$ from interval $0<F\leq1/2$; interval $1/2<F<1$ can be
considered by taking in view that the averages multiplied by $s$
and the fluctuations are invariant under $F\rightarrow1-F$ and
$\Theta\rightarrow-\Theta$. As before, two dashed (with longer and
shorter dashes) and one dotted lines correspond to values
$(k_BT/|m|)=5^{-1}$, 1, 5, and solid lines correspond to values
$T=0,\infty$. In the $F\neq1/2$ case, both fluctuations at
extremely small temperatures possess a peak at $A=-1$ which is
smoothed out as temperature increases; incidentally, the minimum
evolves to the left of point $A^{-1}=0$ in the case of $0<F<1/2$
and to the right of point $A=0$ in the case of $1/2<F<1$. In
contrast to the $\hat R$-fluctuation, the $\hat M$-fluctuation is
obviously negative in the vicinity of this minimum, see Figs.4 and
5. In the $F=1/2$ case, the behaviour of two fluctuation is
completely different, with the $\hat M$-fluctuation being
independent of $\Theta$ and negative, see Fig.6 and explicit
expression \eqref{f1}. The behaviour of the $\hat R$-fluctuation
is qualitatively the same as that of the fermion number
fluctuation studied in detail in Ref.\cite{SiG}. Thus, we conclude
that a physically meaningful observable is the improved total
angular momentum $(\hat R)$ rather than the naive one $(\hat M)$.

At this point it is appropriate to discuss a very tiny effect
which is common for fluctuations of fermion number and improved
total angular momentum. Namely, both fluctuations at rather small
temperatures become negative with extremely small absolute values
in the case of absence of the bound state in the one-particle
spectrum, i.e., at $\cos\Theta>0$ $(A>0)$. Fig.7 illustrates this
fact for the $\hat R$-fluctuation. To be more specific, in the
$F\neq1/2$ case, this fluctuation at $T=|m|/(5k_B)$ attains value
$-10^{-4}$ at $F=0.1$ $(F=0.9)$ and value $-3\times10^{-4}$ at
$F=0.3$ ($F=0.7$) in region $0\,\, ({\rm mod}
2\pi)<\Theta<s\frac\pi2\,\, ({\rm mod} 2\pi)$ $(-s\frac\pi2\,\,
({\rm mod} 2\pi)<\Theta<0\,\, ({\rm mod} 2\pi))$; in the $F=1/2$
case, the fluctuation at $T=|m|/k_B$ attains value
$-27\times10^{-4}$ in region $-s\frac\pi2\,\, ({\rm mod}
2\pi)<\Theta<s\frac\pi2\,\,({\rm mod} 2\pi)$.

Let us recall that boundary parameter $\Theta$ is introduced as a
parameter providing a self-adjoint extension of the Dirac
Hamiltonian in the background of a pointlike vortex defect. Thus
the mathematical requirement of self-adjointness of the
Hamiltonian is consistent with all values of $\Theta$. However,
the physical requirements may restrict the range of $\Theta$. As
we see, this really happens: nonnegativeness of the fluctuations
of fermion number and improved total angular momentum is
consistent with values of $\Theta$ from the range corresponding to
the case of existence of the bound state in the one-particle
spectrum, i.e. $\cos\Theta<0$ $(A<0)$.

To conclude, we reiterate that a comprehensive study of thermal
characteristics of the planar quantum fermionic system with a
pointlike vortex defect has allowed us to reduce the
mathematically acceptable set of boundary conditions at the
location of the defect to the physically acceptable one. It is
found that angular momentum of the system is defined as shifted by
$1/2$ from its naive value.

\section*{Acknowledgements}
This work was partially supported by the State Foundation for
Basic Research of Ukraine (project 2.7/00152).


{ \setcounter{equation}{0}
\renewcommand{\theequation}{A.\arabic{equation}}
\appendix
\section*{Appendix A}

The radial components of the resolvent kernel of $H$ \eqref{c3}
are presented in the following way, type 1 $(l=s(n-n_0)>0)$:
\begin{equation}\label{ap1}
a_n(r;r')=\frac{i\pi}2(\omega+m)\left[\theta(r-r')H^{(1)}_{l-F}(k
r)J_{l-F}(k r')+ \theta(r'-r)J_{l-F}(k r)H^{(1)}_{l-F}(k
r')\right],
\end{equation}
\begin{equation}\label{ap2}
b_n(r;r')=\frac{i\pi}2k\left[\theta(r-r')H^{(1)}_{l+1-F}(k
r)J_{l-F}(k r')+ \theta(r'-r)J_{l+1-F}(k r)H^{(1)}_{l-F}(k
r')\right],
\end{equation}
\begin{equation}\label{ap3}
c_n(r;r')=\frac{i\pi}2(\omega-m)\left[\theta(r-r')H^{(1)}_{l+1-F}(k
r)J_{l+1-F}(k r')+ \theta(r'-r)J_{l+1-F}(k r)H^{(1)}_{l+1-F}(k
r')\right],
\end{equation}
\begin{equation}\label{ap4}
d_n(r;r')=\frac{i\pi}2k\left[\theta(r-r')H^{(1)}_{l-F}(k
r)J_{l+1-F}(k r')+ \theta(r'-r)J_{l-F}(k r)H^{(1)}_{l+1-F}(k
r')\right];
\end{equation}

type 2 $(l'=-s(n-n_0)>0)$:
\begin{equation}\label{ap5}
a_n(r;r')=\frac{i\pi}2(\omega+m)\left[\theta(r-r')H^{(1)}_{l'+F}(k
r)J_{l'+F}(k r')+ \theta(r'-r)J_{l'+F}(k r)H^{(1)}_{l'+F}(k
r')\right],
\end{equation}
\begin{equation}\label{ap6}
b_n(r;r')=-\frac{i\pi}2k\left[\theta(r-r')H^{(1)}_{l'-1+F}(k
r)J_{l'+F}(k r')+ \theta(r'-r)J_{l'-1+F}(k r)H^{(1)}_{l'+F}(k
r')\right],
\end{equation}
\begin{equation}\label{ap7}
c_n(r;r')=\frac{i\pi}2(\omega-m)\left[\theta(r-r')H^{(1)}_{l'-1+F}(k
r)J_{l'-1+F}(k r')+ \theta(r'-r)J_{l'-1+F}(k r)H^{(1)}_{l'-1+F}(k
r')\right],
\end{equation}
\begin{equation}\label{ap8}
d_n(r;r')=-\frac{i\pi}2k\left[\theta(r-r')H^{(1)}_{l'+F}(k
r)J_{l'-1+F}(k r')+ \theta(r'-r)J_{l'+F}(k r)H^{(1)}_{l'-1+F}(k
r')\right];
\end{equation}

type 3 $(n=n_0)$:
\begin{multline}\label{ap9}
a_{n_0}(r;r')=\frac{i\pi}2\frac{\omega+m}{\sin\nu_\omega+\cos\nu_\omega
e^{iF\pi}}\left\{\theta(r-r')H^{(1)}_{-F}(k r) [\sin\nu_\omega
J_{-F}(k r')+\cos\nu_\omega J_{F}(k r')]+\right.\\
+\left.\theta(r'-r)[\sin\nu_\omega J_{-F}(k r)+\cos\nu_\omega
J_{F}(k r)]H^{(1)}_{-F}(k r')
\right\},
\end{multline}
\begin{multline}\label{ap10}
b_{n_0}(r;r')=\frac{i\pi}2\frac{k}{\sin\nu_\omega+\cos\nu_\omega
e^{iF\pi}}\left\{\theta(r-r')H^{(1)}_{1-F}(k r) [\sin\nu_\omega
J_{-F}(k r')+\cos\nu_\omega J_{F}(k r')]+\right.\\
+\left.\theta(r'-r)[\sin\nu_\omega J_{1-F}(k r)-\cos\nu_\omega
J_{-1+F}(k r)]H^{(1)}_{-F}(k r')
\right\},
\end{multline}
\begin{multline}\label{ap11}
c_{n_0}(r;r')=\frac{i\pi}2\frac{\omega-m}{\sin\nu_\omega+\cos\nu_\omega
e^{iF\pi}}\left\{\theta(r-r')H^{(1)}_{1-F}(k r) [\sin\nu_\omega
J_{1-F}(k r')-\cos\nu_\omega J_{-1+F}(k r')]+\right.\\
+\left.\theta(r'-r)[\sin\nu_\omega J_{1-F}(k r)-\cos\nu_\omega
J_{-1+F}(k r)]H^{(1)}_{1-F}(k r')
\right\}\,,
\end{multline}
\begin{multline}\label{ap12}
d_{n_0}(r;r')=\frac{i\pi}2\frac{k}{\sin\nu_\omega+\cos\nu_\omega
e^{iF\pi}}\left\{\theta(r-r')H^{(1)}_{-F}(k r) [\sin\nu_\omega
J_{1-F}(k r')-\cos\nu_\omega J_{-1+F}(k r')]+\right.\\
+\left.\theta(r'-r)[\sin\nu_\omega J_{-F}(k r)+\cos\nu_\omega
J_{F}(k r)]H^{(1)}_{1-F}(k r')
\right\}.
\end{multline}
Here $k=\sqrt{\omega^2-m^2}$ and a physical sheet is chosen as
$0<{\rm Arg}\,k<\pi$ $(Im k>0)$, $\theta(u)=\frac12[1+{\rm
sgn}(u)]$, $J_\rho(u)$ is the Bessel function of order $\rho$,
$H^{(1)}_\rho(u)$ is the first-kind Hankel function of order
$\rho$, and
\begin{equation}\label{ap13}
\tan\nu_\omega=\frac{k^{2F}}{\omega+m}\,{\rm{sgn}}(m)(2|m|)^{1-2F}
\frac{\Gamma(1-F)}{\Gamma(F)}\tan\left(s\frac\Theta2+\frac\pi4\right)\,.
\end{equation}

In the absence of the vortex defect the radial components take
form:
\begin{equation}\label{ap14}
\left.a_n(r;r')\right|_{e\Phi=0}=\frac{i\pi}2(\omega+m)\left[\theta(r-r')H^{(1)}_{n}(k
r) J_{n}(k r')+\theta(r'-r)J_{n}(k r)H^{(1)}_{n}(k r')\right],
\end{equation}
\begin{equation}\label{ap15}
\left.b_n(r;r')\right|_{e\Phi=0}=\frac{i\pi}2k\left[\theta(r-r')H^{(1)}_{n+s}(k
r) J_{n}(k r')+\theta(r'-r)J_{n+s}(k r)H^{(1)}_{n}(k r')\right],
\end{equation}
\begin{equation}\label{ap16}
\left.c_n(r;r')\right|_{e\Phi=0}=\frac{i\pi}2(\omega-m)\left[\theta(r-r')H^{(1)}_{n+s}(k
r) J_{n+s}(k r')+\theta(r'-r)J_{n+s}(k r)H^{(1)}_{n+s}(k
r')\right],
\end{equation}
\begin{equation}\label{ap17}
\left.d_n(r;r')\right|_{e\Phi=0}=\frac{i\pi}2k\left[\theta(r-r')H^{(1)}_{n}(k
r) J_{n+s}(k r')+\theta(r'-r)J_{n}(k r)H^{(1)}_{n+s}(k r')\right].
\end{equation}

Note that all radial components behave asymptotically at large
distances as outgoing waves.


{ \setcounter{equation}{0}
\renewcommand{\theequation}{B.\arabic{equation}}
\section*{Appendix B}
Let us consider quantity \eqref{c11}. The contribution of the
regular (types 1 and 2) components of the resolvent kernel is
given by expression
\begin{multline}\label{bp1}
\sum_{n\neq n_0}[(n-e\Phi)a_n(r;r')+(n+s-e\Phi)c_n(r;r')]=\\
=s\sum_{l\geq 1}\left[(l-F)(\omega+m)I_{l-F}(\kappa
r)K_{l-F}(\kappa r')+(l+1-F)(\omega-m)I_{l+1-F}(\kappa
r)K_{l+1-F}(\kappa r')\right]-\\ -s\sum_{l'\geq
1}\left[(l'+F)(\omega+m)I_{l'+F}(\kappa r)K_{l'+F}(\kappa
r')+(l'-1+F)(\omega-m)I_{l'-1+F}(\kappa r)K_{l'-1+F}(\kappa
r')\right]\,,
\end{multline}
where $\kappa=-ik$ and we used relations
\begin{equation*}
J_\rho(i \kappa r)=e^{i \rho\frac\pi2}I_\rho(\kappa r)\quad {\rm
and} \quad H^{(1)}(i\kappa r')=\frac{2}{i\pi}e^{-i
\rho\frac\pi2}K_\rho(\kappa r)\,,
\end{equation*}
which are valid at $Re\,\kappa>0$; here $I_\rho(u)$ is the
modified Bessel function of order $\rho$, and
\begin{equation*}
 K_\rho(u)=\frac{\pi}{2\sin(\rho\pi)}\left[I_{-\rho}(u)-I_\rho(u)\right]\,.
\end{equation*}
Using relation (see, e.g., Ref.\cite{Prud})
\begin{equation}\label{bp2}
I_\rho(\kappa r)K_\rho(\kappa r')=\frac12\int\limits_0^\infty
\frac{dy}{y}\,\exp\left(-\frac{\kappa^2r
r'}{2y}-\frac{r^2+{r'}^2}{2r r' }\,y\right)I_\rho(y)\,,\quad
Re\,\kappa^2>0\,,
\end{equation}
\begin{equation}\label{bp3}
\sum_{l\geq
1}(l+\rho)I_{l+\rho}(y)=\frac12y\left[I_\rho(y)+I_{\rho+1}(y)\right]\,,
\end{equation}
we perform summation in Eq.\eqref{bp1} and get in the case of
$Re\,\kappa>|Im\,\kappa|$:
\begin{multline}\label{bp4}
\sum_{n\neq n_0}[(n-e\Phi)a_n(r;r')+(n+s-e\Phi)c_n(r;r')]=\\
=\frac{s\sin(F \pi)}{\pi}\omega\int\limits_0^\infty
dy\,\exp\left(-\frac{\kappa^2r r'}{2y}-\frac{r^2+{r'}^2}{2r r'
}\,y\right)\left[K_F(y)-K_{1-F}(y)\right]+\\+s
F(\omega+m)I_F(\kappa r)K_F(\kappa r')-s
(1-F)(\omega-m)I_{1-F}(\kappa r)K_{1-F}(\kappa r')\,.
\end{multline}
The contribution of the irregular (type 3) components is given by
expression
\begin{multline}\label{bp5}
(n_0-e\Phi)a_{n_0}(r;r')+(n_0+s-e\Phi)c_{n_0}(r;r')
=\frac{2s\sin(F\pi)}{\pi(\tan\nu_\omega+e^{i F\pi})}\times\\
\times\left[-F(\omega+m)\tan\nu_\omega K_F(\kappa r)K_F(\kappa
r')+(1-F)(\omega-m)e^{i F\pi}K_{1-F}(\kappa r)K_{1-F}(\kappa
r')\right]-\\ -s F(\omega+m)I_F(\kappa r)K_F(\kappa r')+s
(1-F)(\omega-m)I_{1-F}(\kappa r)K_{1-F}(\kappa r')\,.
\end{multline}
Summing Eqs.\eqref{bp4} and \eqref{bp5}, we get Eq.\eqref{c17}.

Computation of other quantities, Eqs.\eqref{c12}-\eqref{c16}, is
similar to the above. As an illustration, let us scrutinize the
procedure of computation of the last quantity, Eq.\eqref{c16},
which is the most tedious one.

The contribution of the regular components is
\begin{multline}\label{bp6}
\frac{e^2}{4\pi}\,sr\sum_{n\neq
n_0}\left(n-e\Phi+\frac12s\right)\left[b_n(r;r')+d_n(r;r')\right]=\\
 =-\frac{e^2}{4\pi}\kappa r \left\{\sum_{l\geq1}\left(l-F+\frac12\right)
\left[I_{l+1-F}(\kappa r)K_{l-F}(\kappa r')-I_{l-F}(\kappa
r)K_{l+1-F}(\kappa r')\right]-\right.\\
-\left.\sum_{l'\geq1}\left(l'+F-\frac12\right)
\left[I_{l'-1+F}(\kappa r)K_{l'+F}(\kappa r')-I_{l'+F}(\kappa
r)K_{l'-1+F}(\kappa r')\right]\right\}.
\end{multline}
Using recurrency relation
\begin{equation*}
u\partial_u I_\rho(u)=\pm \rho I_\rho(u)+u I_{\rho\pm1}(u),
\end{equation*}
we get
\begin{multline*}
\frac{e^2}{4\pi}s r\sum_{n\neq
n_0}\left(n-e\Phi+\frac12s\right)\left[b_n(r;r')+d_n(r;r')\right]=\\
 =\frac{e^2}{4\pi}\sum_{l\geq1}\left\{
\left(l-F+\frac12\right)\left[(l-F)I_{l-F}(\kappa r)K_{l-F}(\kappa
r')+(l+1-F)I_{l+1-F}(\kappa r)K_{l+1-F}(\kappa
r')\right]+\right.\\ +\left.\left(l+F-\frac12\right)
\left[(l+F)I_{l+F}(\kappa r)K_{l+F}(\kappa
r')+(l-1+F)I_{l-1+F}(\kappa r)K_{l-1+F}(\kappa
r')\right]\right\}-\\-\frac{e^2}{4\pi}r\partial_r\sum_{l\geq1}\left\{
\left(l-F+\frac12\right)\left[I_{l-F}(\kappa r)K_{l-F}(\kappa
r')-I_{l+1-F}(\kappa r)K_{l+1-F}(\kappa r')\right]-\right.\\
-\left.\left(l+F-\frac12\right) \left[I_{l+F}(\kappa
r)K_{l+F}(\kappa r')-I_{l-1+F}(\kappa r)K_{l-1+F}(\kappa
r')\right]\right\}=
\end{multline*}
\begin{multline}\label{bp7}
 =\frac{e^2}{16\pi}\sum_{l\geq1}\int\limits_0^\infty
dy\,\exp\left(-\frac{\kappa^2r r'}{2y}-\frac{r^2+{r'}^2}{2r r'
}\,y\right)\left\{\frac32I_{l-1-F}(y)+\frac12I_{l+1-F}(y)+\frac12I_{l-F}(y)+\right.\\+
\frac32I_{l+2-F}(y)+\frac12I_{l-1+F}(y)+\frac32I_{l+1+F}(y)+\frac32I_{l-2+F}(y)+\frac12I_{l+F}(y)+\\
+(l-1-F)I_{l-1-F}(y)-(l+1-F)I_{l+1-F}(y)+(l-F)I_{l-F}(y)-(l+2-F)I_{l+2-F}(y)+\\
+(l-1+F)I_{l-1+F}(y)-(l+1+F)I_{l+1+F}(y)+(l-2+F)I_{l-2+F}(y)-(l+F)I_{l+F}(y)+\\
+\left(\frac{\kappa^2r r'}{y^2}+\frac{r^2-{r'}^2}{r
r'}\right)\left[(1-F)I_{1-F}(y)+F\,I_F(y)+\frac12I_{l-F}(y)+\frac12
I_{l+1-F}(y)+\right.\\+\left.\left.\frac12I_{l+F}(y)+I_{l-1+F}(y)\right]\right\},
\end{multline}
where the second equality is obtained with the use of
representation \eqref{bp2} and recurrency relation
\begin{equation*}
2\rho I_\rho(u)=u\left[I_{\rho-1}(u)-I_{\rho+1}(u)\right].
\end{equation*}
Now the summation over $l$ can be performed with the use of
Eq.\eqref{bp3} and relation (see Ref.\cite{Prud})
\begin{equation}\label{bp8}
\sum_{l=1}^\infty
I_{l+\rho}(y)=-\frac2\rho\left[e^y\int\limits_0^y
du\,e^{-u}I_\rho(u)-yI_{\rho}(y)-yI_{\rho+1}(y)\right],\quad
Re\,\rho>-1.
\end{equation}
Using again the recurrency relation and the relation between the
Macdonald and the modified Bessel functions, we get
\begin{multline}\label{bp9}
\frac{e^2}{4\pi}s r\sum_{n\neq
n_0}\left(n-e\Phi+\frac12s\right)\left[b_n(r;r')+d_n(r;r')\right]=\\
=\frac{e^2\sin(F \pi)}{8F(1-F)\pi^2}\int\limits_0^\infty\!\!
dy\,\left(1+\frac{\kappa^2r r'}{4y^2}\!+\!\frac{r^2-{r'}^2}{4r r'
}\right)\exp\left(\!\!-\frac{\kappa^2r
r'}{2y}\!-\!\frac{r^2+{r'}^2}{2r r' }\,y\!\right)\times \\ \times
\left\{e^y\int\limits_0^y du\,e^{-u}[(1-F)K_F(u)+F
K_{1-F}(u)]+(2F-1)y[K_F(y)-K_{1-F}(y)]\right\}+\\+
\frac{e^2}{8\pi}\int\limits_0^\infty
dy\,\exp\left(-\frac{\kappa^2r r'}{2y}-\frac{r^2+{r'}^2}{2r r'
}\,y\right)\left\{\frac{2\sin(F
\pi)}{\pi}\left[(1-F)K_F(y)+F\,K_{1-F}(y)\right]\right.-\\
-\left.\frac1y
\left(\!F-\frac12\right)\![F\,I_F(y)-(1-F)I_{1-F}(y)]+\frac12
\left(\frac{\kappa^2r r'}{y^2}\!+\!\frac{r^2-{r'}^2}{r r'
}\right)\!\!\left(\!F-\frac12\right)\![I_F(y)-I_{1-F}(y)]
\!\right\}\!.
\end{multline}

The contribution of the irregular components is
\begin{multline}\label{bp10}
\frac{e^2}{4\pi}s
r\left(n_0-e\Phi+\frac12s\right)\left[b_{n_0}(r;r')+d_{n_0}(r;r')\right]=
-\frac{e^2\left(F-\frac12\right)\sin(F\pi)}{2\pi^2(\tan\nu_\omega+e^{i
F\pi})}\,\kappa r\times\\\times\left[\tan\nu_\omega K_{1-F}(\kappa
r)K_F(\kappa r')-e^{i F\pi}K_{F}(\kappa r)K_{1-F}(\kappa
r')\right]+\\+ \frac{e^2}{4\pi}\left(F-\frac12\right)\kappa
r\,[I_{-1+F}(\kappa r)K_F(\kappa r')-I_{-F}(\kappa
r)K_{1-F}(\kappa r')].
\end{multline}

The terms containing the $I_\rho$-functions are cancelled in the
sum of Eqs.\eqref{bp9} and \eqref{bp10}. Decomposing the integral
over $u$ as
$\int\limits_0^y=\int\limits_0^\infty-\int\limits_y^\infty$, we
get the following expression for this sum:
\begin{multline}\label{bp11}
\frac{e^2}{4\pi}s r\sum_{n=-\infty
}^\infty\left(n-e\Phi+\frac12s\right)\left[b_n(r;r')+d_n(r;r')\right]=\\=
\frac{e^2}{4\pi}\int\limits_0^\infty dy\left(1+\frac{\kappa^2r
r'}{4y^2}+\frac{r^2-{r'}^2}{4r r'
}\right)\exp\left(-\frac{\kappa^2r r'}{2y}-\frac{(r-r')^2}{2r r'
}\,y\right)-\\- \frac{e^2\sin(F
\pi)}{8F(1-F)\pi^2}\int\limits_0^\infty
dy\,\left(1+\frac{\kappa^2r r'}{4y^2}+\frac{r^2-{r'}^2}{4r r'
}\right)\exp\left(-\frac{\kappa^2r r'}{2y}-\frac{r^2+{r'}^2}{2r r'
}\,y\right)\times \\ \times \left\{e^y\int\limits_y^\infty
du\,e^{-u}[(1-F)K_F(u)+F
K_{1-F}(u)]-(2F-1)y[K_F(y)-K_{1-F}(y)]\right\}+\\+ \frac{e^2\sin(F
\pi)}{4\pi^2}\int\limits_0^\infty dy\,\exp\left(-\frac{\kappa^2r
r'}{2y}-\frac{r^2+{r'}^2}{2r r'
}\,y\right)\left[(1-F)K_F(y)+F\,K_{1-F}(y)\right]-\\-\frac{e^2\left(F-\frac12\right)\sin(F\pi)}{2\pi^2(\tan\nu_\omega+e^{i
F\pi})}\,\kappa r\left[\tan\nu_\omega K_{1-F}(\kappa r)K_F(\kappa
r')-e^{i F\pi}K_{F}(\kappa r)K_{1-F}(\kappa r')\right].
\end{multline}
Using relation (see, e.g., Ref.\cite{Prud2})
\begin{equation*}
\int\limits_0^\infty dy\,y^{s-1}\exp(-py-qy^{-1})=2\left(\frac q
p\right)^{\frac s2}K_s(2\sqrt{pq}),
\end{equation*}
we express the first integral over $y$ in Eq.\eqref{bp11} through
the Macdonald function and get Eq.\eqref{c22}.

In the case of the absence of the vortex defect one uses summation
formulae
\begin{equation*}
\sum_{n=-\infty}^\infty I_n(y)=e^y,\quad \sum_{n=-\infty}^\infty n
I_n(y)=0.
\end{equation*}
Thus, in particular, we get
\begin{multline}\label{bp12}
\left.\frac{e^2}{4\pi}s r\sum_{n=-\infty
}^\infty\left(n+\frac12s\right)\left[b_n(r;r')+d_n(r;r')\right]\right|_{e\Phi=0}=\\=
-\frac{e^2}{4\pi}\kappa
r\sum_{n=-\infty}^\infty\left(sn+\frac12\right)[I_{sn+1}(\kappa
r)K_{sn}(\kappa r')-I_{sn}(\kappa r)K_{sn+1}(\kappa
r')]=\\=\frac{e^2}{4\pi}\sum_{n=-\infty}^\infty\left(sn+\frac12\right)
\{sn I_{sn}(\kappa r)K_{sn}(\kappa r')+(sn+1)I_{sn+1}(\kappa
r)K_{sn+1}(\kappa r')-\\-r\partial_r[ I_{sn}(\kappa
r)K_{sn}(\kappa r')-I_{sn+1}(\kappa r)K_{sn+1}(\kappa r')]\}=\\
=\frac{e^2}{16\pi}\sum_{n=-\infty}^\infty \int\limits_0^\infty
dy\,\exp\left(-\frac{\kappa^2r r'}{2y}-\frac{r^2+{r'}^2}{2r r'
}\,y\right)\left(sn+\frac12\right)\times\\\times\{I_{sn-1}(y)-I_{sn+1}(y)+I_{sn}(y)-I_{sn+2}(y)+
\left(\frac{\kappa^2r r'}{y^2}+\frac{r^2-{r'}^2}{r
r'}\right)[I_{sn}(y)-I_{sn+1}(y)]\}=\\=
\frac{e^2}{4\pi}\int\limits_0^\infty dy\,\left(1+\frac{\kappa^2r
r'}{4y^2}+\frac{r^2-{r'}^2}{4r r'
}\right)\exp\left(-\frac{\kappa^2r r'}{2y}-\frac{(r-r')^2}{2r r'
}\,y\right),
\end{multline}
which is Eq.\eqref{c26}.



\begin{figure}
\begin{tabular}{c}
\!\!\includegraphics[width=140mm]{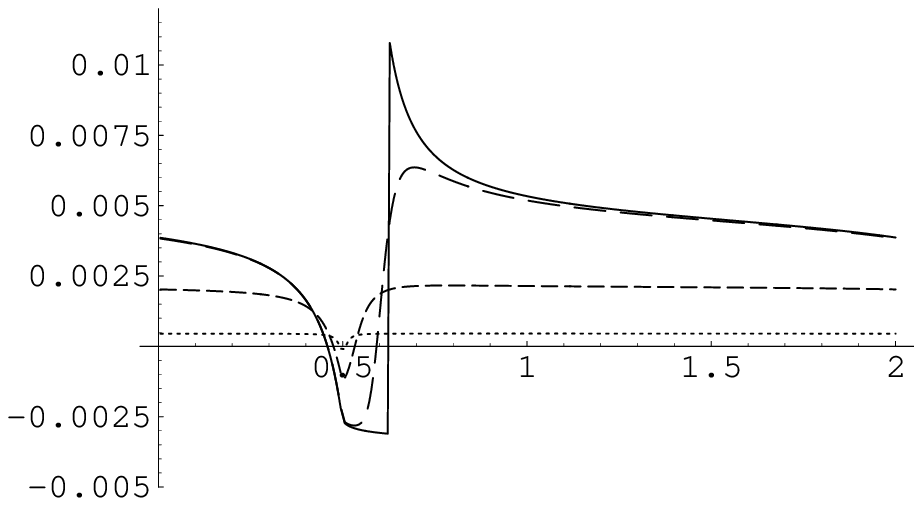}
\put(-320,225){$s|m|e^{-2}O(T)$}\put(-35,58){$s\Theta\pi^{-1}$}
\end{tabular}
\begin{tabular}{c}
\includegraphics[width=140mm]{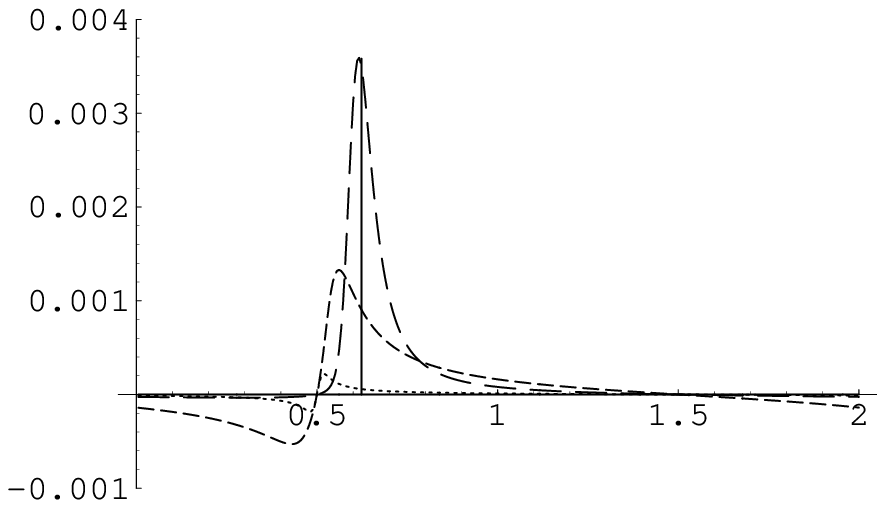}
\end{tabular}
\put(-333,110){$sme^{-2}\Delta(T;\hat O,\hat
N)$}\put(-40,-83){$s\Theta\pi^{-1}$} \caption{$F=0.1$}\label{1}
\end{figure}

 \clearpage
\begin{figure}
\begin{tabular}{c}
\,\,\,\,\,\,\includegraphics[width=140mm]{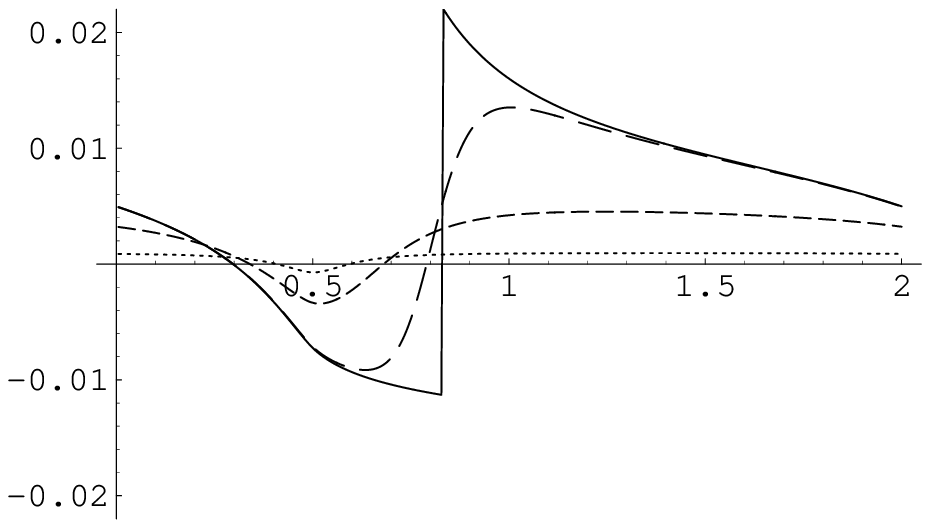}
\put(-340,230){$s|m|e^{-2}O(T)$}\put(-35,100){$s\Theta\pi^{-1}$}
\end{tabular}
\begin{tabular}{c}
\includegraphics[width=140mm]{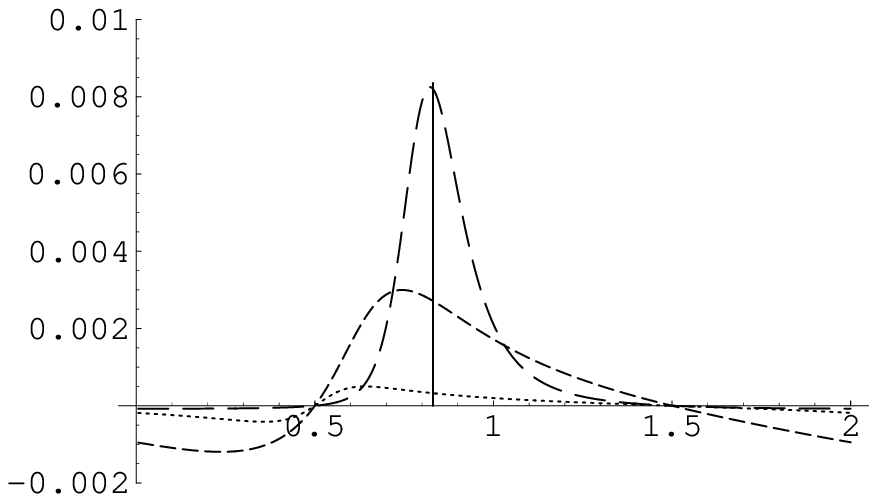}
\end{tabular}
\put(-333,110){$sme^{-2}\Delta(T;\hat O,\hat
N)$}\put(-40,-58){$s\Theta\pi^{-1}$} \caption{$F=0.3$}\label{2}
\end{figure}

\clearpage
\begin{figure}
\begin{tabular}{c}
\,\,\,\,\,\,\includegraphics[width=140mm]{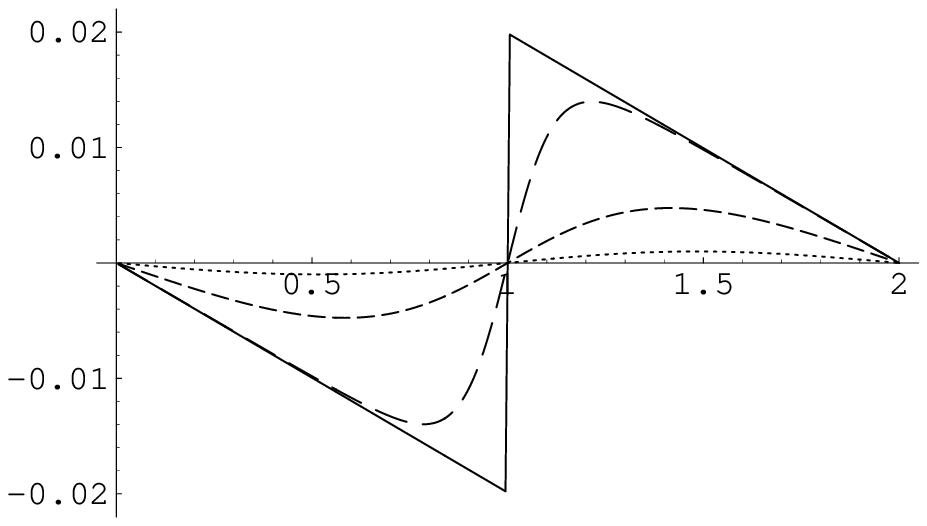}
\put(-340,230){$s|m|e^{-2}O(T)$}\put(-35,98){$s\Theta\pi^{-1}$}
\end{tabular}
\begin{tabular}{c}
\includegraphics[width=140mm]{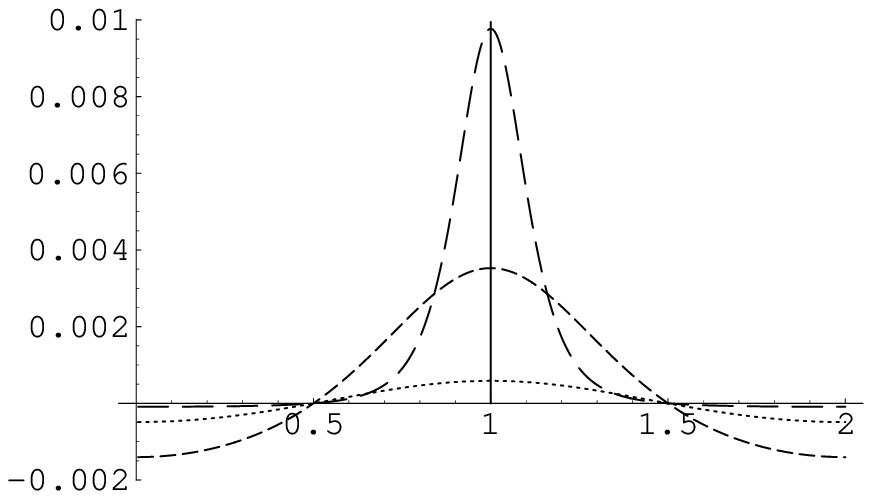}
\end{tabular}
\put(-333,110){$sme^{-2}\Delta(T;\hat O,\hat
N)$}\put(-40,-60){$s\Theta\pi^{-1}$} \caption{$F=0.5$}\label{3}
\end{figure}

\clearpage
\begin{figure}[t]
\begin{tabular}{l}
\includegraphics[width=70mm]{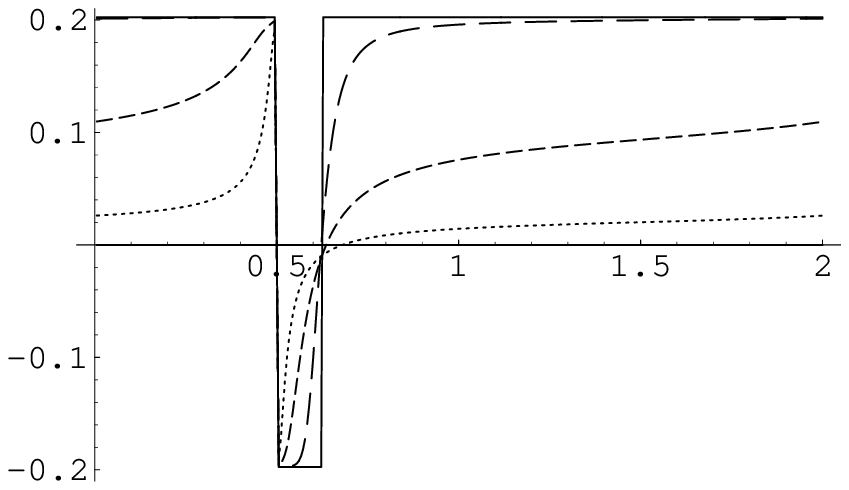}
\end{tabular}
 \put(-175,66){$s\,{\rm{sgn}}(m)M(T)$}
\put(-40,-14){$s\Theta\pi^{-1}$}
\begin{tabular}{r}
\includegraphics[width=70mm]{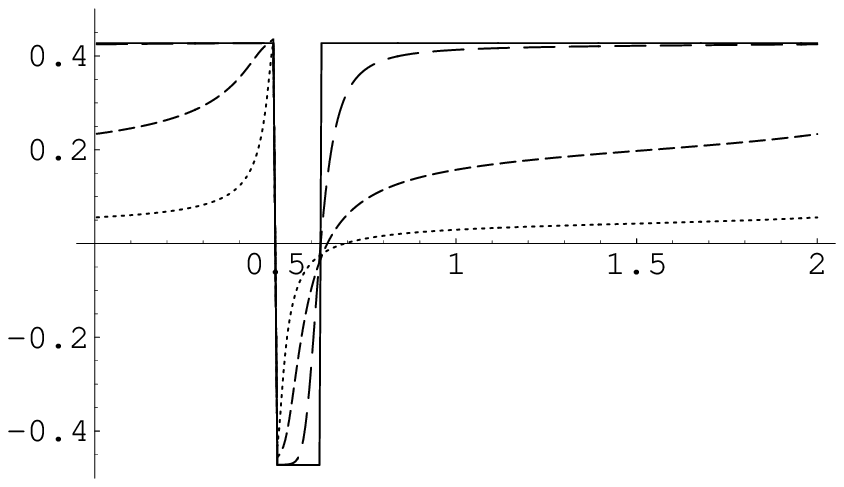}
\end{tabular}
\put(-175,66){$s\,{\rm{sgn}}(m)R(T)$}\put(-40,-14){$s\Theta\pi^{-1}$}
\end{figure}

\begin{figure}[h]
\begin{tabular}{l}
\includegraphics[width=70mm]{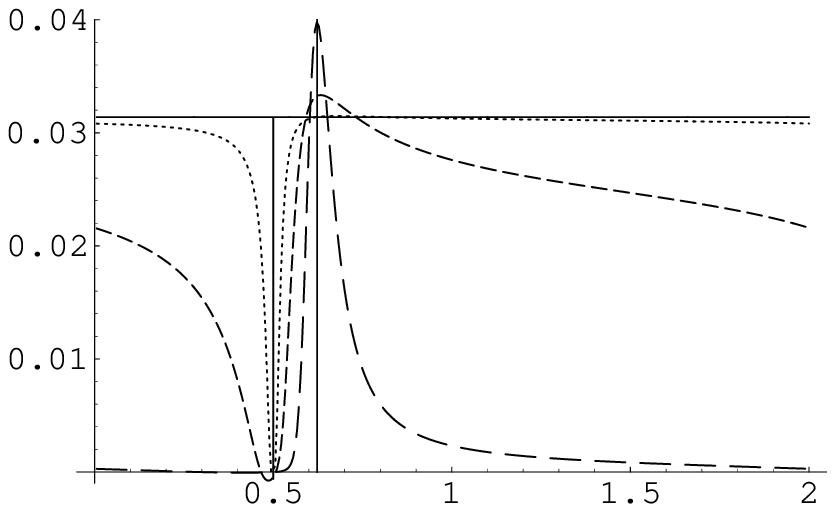}
\end{tabular}
\put(-175,66){$\Delta(T;\hat M,\hat
M)$}\put(-40,-38){$s\Theta\pi^{-1}$}
\begin{tabular}{r}
\includegraphics[width=70mm]{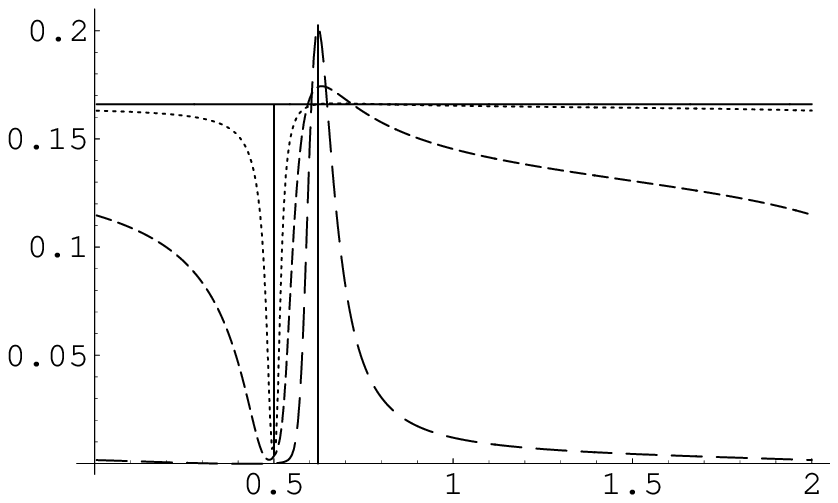}
\end{tabular}
\put(-175,66){$\Delta(T;\hat R,\hat
R)$}\put(-40,-38){$s\Theta\pi^{-1}$} \caption{$F=0.1$}
\end{figure}

\clearpage
\begin{figure}[t]
\begin{tabular}{l}
\,\,\,\,\,\includegraphics[width=70mm]{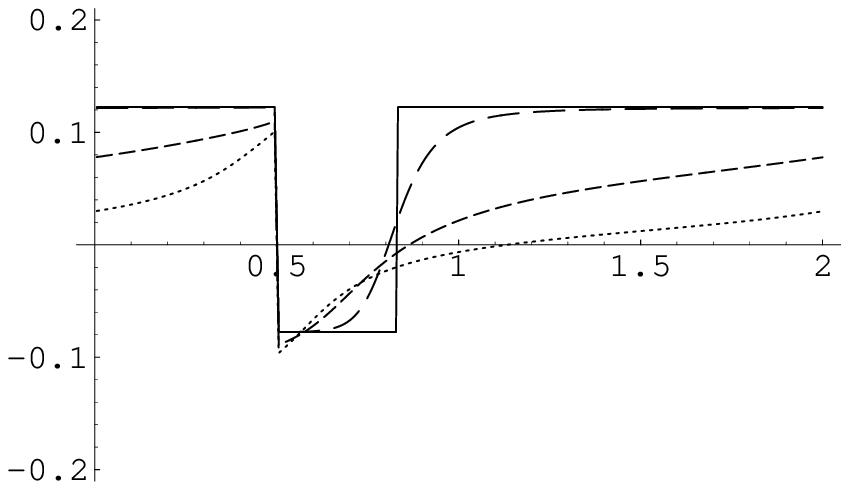}
\end{tabular}
 \put(-175,63){$s\,{\rm{sgn}}(m)M(T)$}
\put(-40,-14){$s\Theta\pi^{-1}$}
\begin{tabular}{r}
\includegraphics[width=70mm]{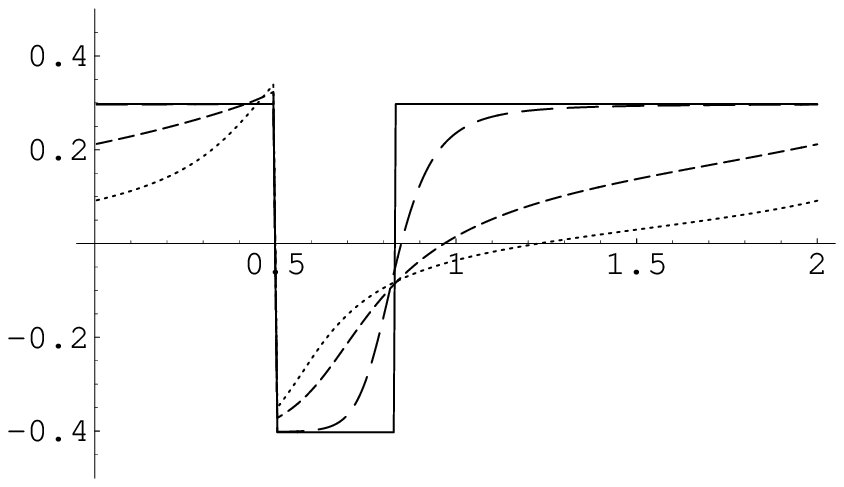}
\end{tabular}
\put(-175,63){$s\,{\rm{sgn}}(m)R(T)$}\put(-40,-14){$s\Theta\pi^{-1}$}
\end{figure}

\begin{figure}[h]
\begin{tabular}{l}
\includegraphics[width=70mm]{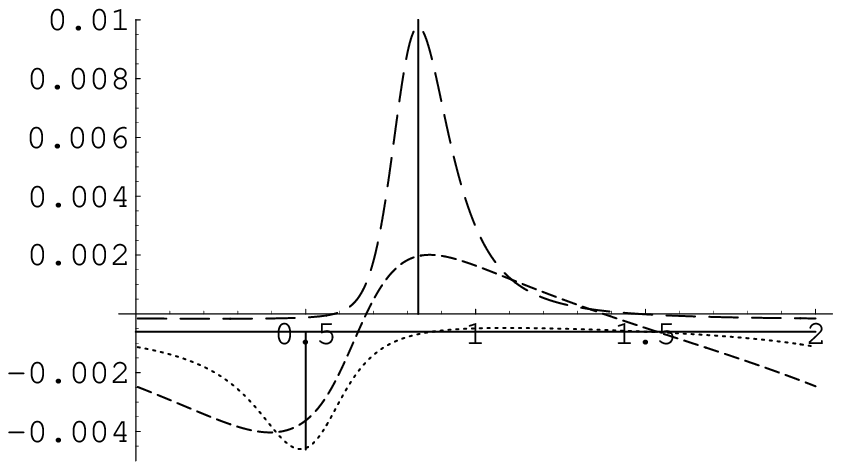}
\end{tabular}
\put(-165,63){$\Delta(T;\hat M,\hat
M)$}\put(-40,-9){$s\Theta\pi^{-1}$}
\begin{tabular}{r}
\,\,\,\,\,\includegraphics[width=70mm]{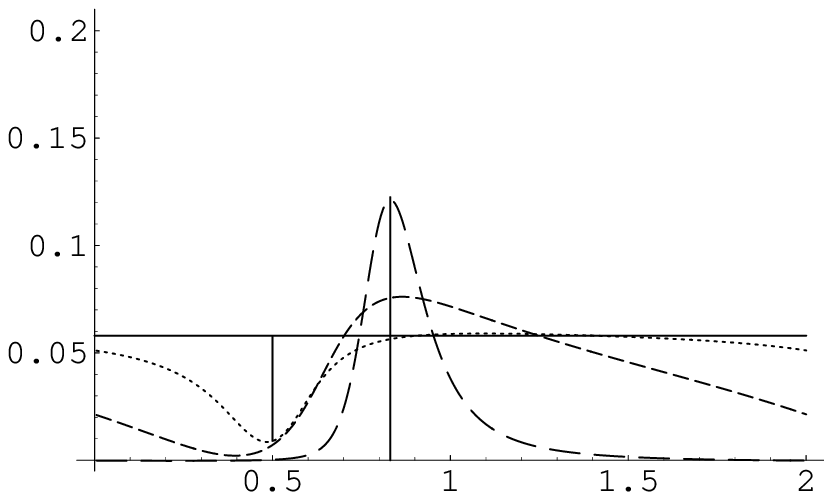}
\end{tabular}
\put(-175,66){$\Delta(T;\hat R,\hat
R)$}\put(-40,-63){$s\Theta\pi^{-1}$} \caption{$F=0.3$}
\end{figure}

\clearpage
\begin{figure}[t]
\begin{tabular}{l}
\,\,\,\,\includegraphics[width=70mm]{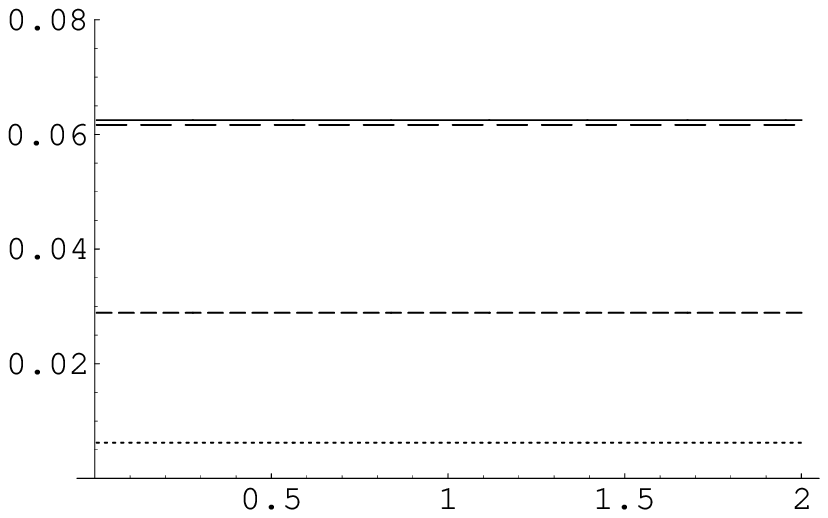}
\end{tabular}
 \put(-175,63){$s\,{\rm{sgn}}(m)M(T)$}
\put(-42,-65){$s\Theta\pi^{-1}$}
\begin{tabular}{r}
\includegraphics[width=70mm]{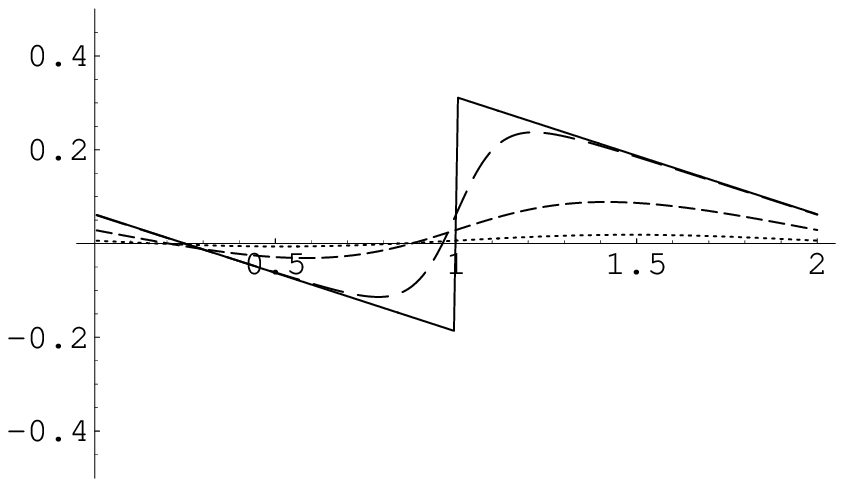}
\end{tabular}
\put(-175,63){$s\,{\rm{sgn}}(m)R(T)$}\put(-40,-14){$s\Theta\pi^{-1}$}
\end{figure}

\begin{figure}[h]
\begin{tabular}{l}
\includegraphics[width=70mm]{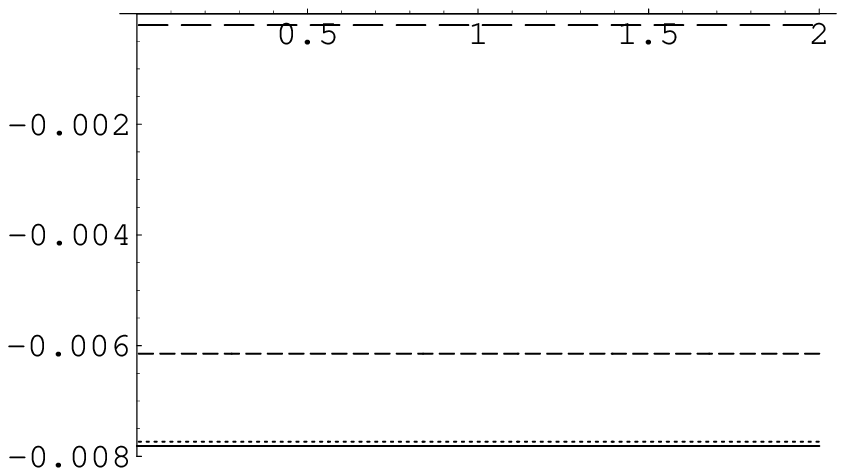}
\end{tabular}
\put(-175,66){$\Delta(T;\hat M,\hat
M)$}\put(-40,65){$s\Theta\pi^{-1}$}
\begin{tabular}{r}
\,\,\,\,\includegraphics[width=70mm]{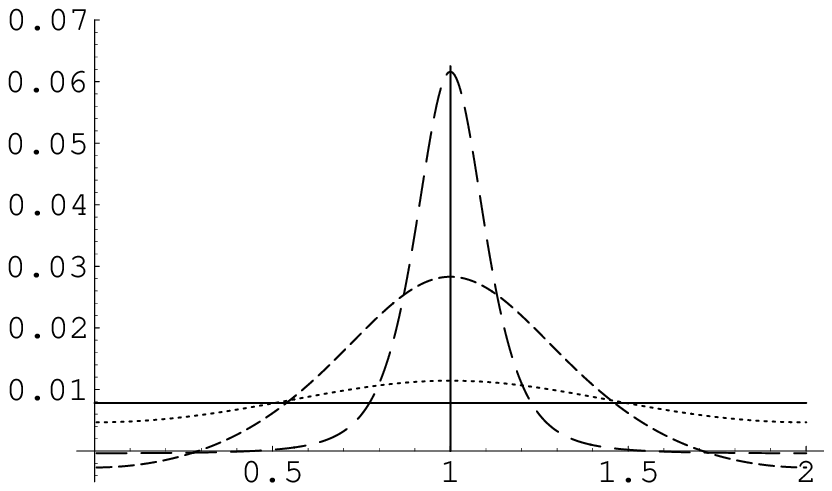}
\end{tabular}
\put(-175,63){$\Delta(T;\hat R,\hat
R)$}\put(-40,-63){$s\Theta\pi^{-1}$} \caption{$F=0.5$}
\end{figure}

\clearpage
\begin{figure}[t]
\begin{tabular}{c}
\,\,\includegraphics[width=80mm]{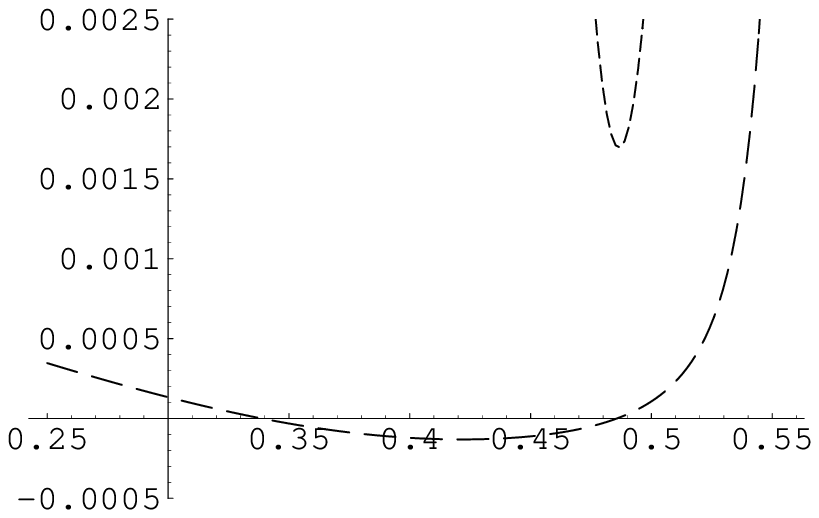}
\end{tabular}
\put(-25,-25){a)} \put(-280,65){$\Delta(T;\hat R,\hat R)$}
\put(-40,-60){$s\Theta\pi^{-1}$}
\end{figure}
\begin{figure}[h]
\begin{tabular}{c}
\,\,\,\,\,\,\includegraphics[width=80mm]{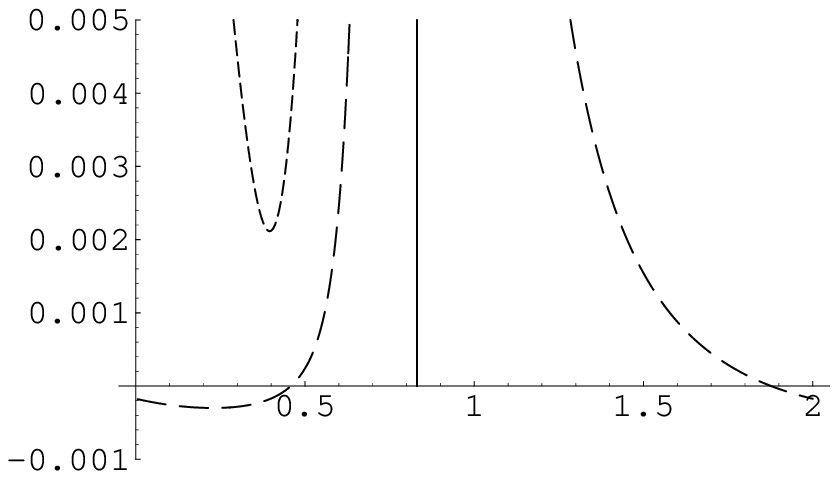}
\end{tabular}
\put(-25,-22){b)} \put(-285,60){$\Delta(T;\hat R,\hat
R)$}\put(-40,-56){$s\Theta\pi^{-1}$}
\end{figure}
\begin{figure}[h]
\begin{tabular}{c}
\,\,\,\,\,\,\includegraphics[width=80mm]{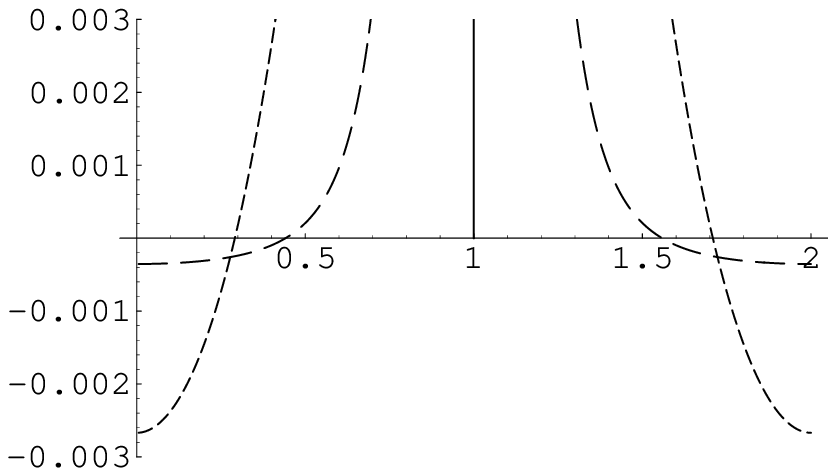}
\end{tabular}
\put(-25,57){c)} \put(-285,60){$\Delta(T;\hat R,\hat
R)$}\put(-36,10){$s\Theta\pi^{-1}$} \caption{a) $F=0.1$, b)
$F=0.3$, c) $F=0.5$}
\end{figure}


\begin {thebibliography}{99}
\raggedright
\bibitem{1} R. Jackiw, C. Rebbi, Phys. Rev. D\textbf{13} (1976) 3398.
\bibitem{2} J. Goldstone, F. Wilczek, Phys. Rev. Lett. \textbf{47} (1981) 986.
\bibitem{3} A. J. Niemi, G. W. Semenoff, Phys. Rep. \textbf{135} (1986) 99.
\bibitem{4} F. Wilczek, cond-mat/0206122.
\bibitem{5} A. J. Niemi, G.W. Semenoff, Phys. Lett. B\textbf{135} (1984) 121;
            A. J. Niemi, Nucl. Phys.  B{\bf251} [FS13]  (1985) 155.
\bibitem{Cor} C. Coriano, R. Parwani,  Phys. Lett. B\textbf{363} (1995) 71;
              A. Goldhaber, R. Parwani, H. Singh, Phys. Lett. B\textbf{386} (1996) 207.
\bibitem{Du0}  G. V. Dunne, J. Feinberg, Phys. Lett. B\textbf{477} (2000) 474.
\bibitem{Du1}  I. J. Aitchison, G. V. Dunne, Phys. Rev. Lett. \textbf{86} (2001) 1690;
               G. V. Dunne, K. Rao, Phys. Rev. D\textbf{64} (2001) 025003;
               G. V. Dunne, Nucl. Phys. Proc. Suppl. \textbf{108} (2002) 155.
\bibitem{Du2} G. V. Dunne, J. Lopez-Sarrion, K. Rao, Phys. Rev. D\textbf{66} (2002) 025004;
               G. V. Dunne, K. Rao, Phys. Rev. D\textbf{67} (2003) 045013.
\bibitem{Vol} G. E. Volovik, JETP Lett. \textbf{70}, (1999) 792;
              Proc. Nat. Acad. Sci. 97 (2000) 2431.
\bibitem{BaE} E. Babaev, Phys. Rev. Lett. \textbf{89} (2002) 067001.
\bibitem{Fra} M. Franz, Z. Tesanovic, O. Vafek, Phys. Rev. B\textbf{66} (2002) 054535.
\bibitem{Semenr} G. W. Semenoff, Phys. Rev. D\textbf{37} (1988) 2838.
\bibitem{Forter} S. Forte, Phys. Rev. D\textbf{38} (1988) 1108.
\bibitem{Aha} Y. Aharonov, D. Bohm, Phys. Rev. \textbf{115} (1959) 485.
\bibitem{Ser} E. M. Serebryanyi, Theor. Math. Phys. 64 (1985) 846.
\bibitem{Si8} Yu. A. Sitenko, Sov. J. Nucl. Phys. \textbf{47} (1988)
              184; \textbf{48} (1988) 670.
\bibitem{Si0} Yu. A. Sitenko, Nucl. Phys. B\textbf{342} (1990) 655;
              Phys. Lett. B\textbf{253} (1991) 138.
\bibitem{Gor} P. Gornicki, Ann. Phys. (N.Y.) \textbf{202} (1990) 271;
              E. G. Flekkoy, J. M. Leinaas, Int. J. Mod. Phys. A\textbf{6} (1991) 5327.
\bibitem{SiR} Yu. A. Sitenko, D. G. Rakityansky, Ukrain. J. Phys.
              \textbf{41} (1996) 329; Phys. Atom. Nucl. \textbf{60} (1997) 247; 258.
\bibitem{Si6} Yu. A. Sitenko, Phys. Lett. B\textbf{387} (1996) 334;
               Yu. A. Sitenko, D. G. Rakityansky, Phys. Atom. Nucl. \textbf{60} (1997) 1497.
\bibitem{Si7} Yu. A. Sitenko, Phys. Atom. Nucl. \textbf{60} (1997) 2102;
              (E) \textbf{62} (1999) 1084.
\bibitem{Si9}  Yu. A. Sitenko, Phys. Atom. Nucl. \textbf{62} (1999) 1056; 1767.
\bibitem{SiG} Yu. A. Sitenko, V. M. Gorkavenko, Nucl. Phys. B\textbf{679} [FS] (2004) 597.
\bibitem{Paranr} M. Paranjape,  Phys. Rev. Lett. \textbf{55} (1985) 2390;
                 Phys. Rev. D\textbf{36} (1987) 3766.
\bibitem{Wil82} F. Wilczek, Phys. Rev. Lett. \textbf{48} (1982) 1144; \textbf{49} (1982) 957.
\bibitem{Jac83} R. Jackiw, A. N. Redlich, Phys. Rev. Lett. \textbf{50} (1983) 555.
\bibitem{Alb} S. Albeverio, F. Gesztezy, R. Hoegh-Krohn, H. Holden,
              \textit{Solvable models in Quantum Mechanics}, Springer-Verlag, Berlin, 1988.
\bibitem{Cas} H. B. G. Casimir, Proc. Kon. Ned. Akad. Wetenschap B \textbf{51}
(1948) 793; Physica (Amsterdam) \textbf{19} (1953) 846.
\bibitem{Prud} A. P. Prudnikov, Yu. A. Brychkov, O. I. Marychev,
               {\em Integrals and Series: Special Functions}, Gordon \& Breach, New York, 1986.
\bibitem{Prud2} A. P. Prudnikov, Yu. A. Brychkov, O. I. Marychev,
               {\em Integrals and Series: Elementary Functions}, Gordon \& Breach, New York, 1982.

\end{thebibliography}
\end{document}